\documentclass[11pt,a4paper]{article}
\pdfoutput=1
\usepackage{jheppub,bm,booktabs,multirow}
\usepackage{color}
\usepackage{overpic}
\allowdisplaybreaks

\makeatletter
\def\@fpheader{~}
\makeatother

\usepackage[Q=yes,pverb-linebreak=no]{examplep}

\def\e{\epsilon}
\def\nno{\nonumber}

\preprint{\begin{flushright}
CERN-TH-2018-187
\end{flushright}}

\title{Non-global logarithms in jet and isolation cone cross sections}
\author[a]{Marcel Balsiger}
\author[a]{\!, Thomas Becher}
\author[b]{and Ding Yu Shao}

\affiliation[a]{Albert Einstein Center for Fundamental Physics, Institut f\"ur Theoretische Physik, Universit\"at Bern,
  Sidlerstrasse 5, CH-3012 Bern, Switzerland}
  \affiliation[b]{CERN, Theoretical Physics Department, CH-1211, Geneva 23, Switzerland}

\emailAdd{balsiger@itp.unibe.ch}
\emailAdd{becher@itp.unibe.ch}
\emailAdd{dingyu.shao@cern.ch}

\date{\today}

\abstract{Starting from a factorization theorem in effective field theory, we derive a parton-shower equation for the resummation of non-global logarithms.  We have implemented this shower and interfaced it with a tree-level event generator to obtain an automated framework to resum the leading logarithm of non-global observables in the large-$N_c$ limit. Using this setup, we compute gap fractions for dijet processes and isolation cone cross sections relevant for photon production. We compare our results with fixed-order computations and LHC measurements. We find that naive exponentiation is often not adequate, especially when the vetoed region is small, since non-global contributions are enhanced due to their dependence on the veto-region size. Since our parton shower is derived from first principles and based on renormalization-group evolution, it is clear what ingredients will have to be included to perform resummations at subleading logarithmic accuracy in the future.}

\begin{document}

\maketitle

\section{Introduction}

In the papers \cite{Becher:2015hka,Becher:2016mmh} we have derived a factorization formula for exclusive jet cross sections which allows one to resum the logarithms arising in the limit where the energy $Q_0$ outside the jets is much smaller than the energy $Q$ inside. In these papers, we have computed different ingredients of the factorization theorem and verified that the logarithmic structure is fully reproduced at Next-to-Next-to-Leading Order (NNLO), but no resummation was performed. Also, for simplicity, we focussed on the Sterman-Weinberg jet cross section, which is defined for  $e^+e^-$ colliders. In the present paper we follow up on the work \cite{Becher:2015hka,Becher:2016mmh} and discuss the resummation of the leading non-global logarithms (NGLs) in detail. We show that the renormalization group (RG) equation which drives it translates into a parton-shower equation. Implementing this shower then allows us to resum a variety of non-global observables.

That the complicated pattern of logarithms for non-global observables can be obtained from an angular dipole shower was shown already in the original paper by Dasgupta and Salam who discovered them \cite{Dasgupta:2001sh}. Their analysis was based on the properties of strongly-ordered QCD amplitudes. The connection to parton showers is less immediate in our treatment which is based on RG evolution in Soft-Collinear Effective Theory (SCET) \cite{Bauer:2000yr,Bauer:2001yt,Beneke:2002ph} (see \cite{Becher:2014oda,Becher:2018gno} for a review). Our starting point is a factorization theorem which separates the hard radiation inside the jets (or outside the isolation cone) from the soft radiation. The soft radiation is driven by Wilson lines along the directions of the hard partons in the process. Since there are contributions involving any number of hard partons, we end up with operators with an arbitrary number of Wilson lines and these operators mix under renormalization. The corresponding RG equation is complicated, but we will show that it takes the form of a recursive equation that can be solved using a parton-shower Monte Carlo (MC) program, which at leading-log accuracy in the large-$N_c$ limit is equivalent to the one used by Dasgupta and Salam. An advantage of our treatment is that the RG equation is not limited to leading logarithmic accuracy and we briefly discuss which ingredients and modifications will be necessary to reach higher precision. There has been a lot of recent work \cite{Li:2016yez,Nagy:2017ggp,Hoche:2017iem,Hoche:2017hno} on the general structure of parton showers and how to increase their accuracy. The problem at hand provides an explicit example of a shower equation derived from first principles for which it is clear what ingredients are needed to resum sub-leading logarithms.

The leading logarithms can be obtained by starting from the tree-level amplitudes and running the parton shower to generate the logarithmically enhanced terms. We have written a dedicated parton-shower code to perform the resummation and use the {\sc MadGraph5\Q{_}aMC@NLO} framework \cite{Alwall:2014hca} to generate the necessary tree-level amplitudes. This provides an automated framework to perform the resummation, which we use to study exclusive jet and isolation cone cross sections, both at lepton and hadron colliders. In particular, we give numerical results for dijet production with a gap between jets and compare to ATLAS measurements \cite{Aad:2011jz,Aad:2014pua} and theoretical predictions \cite{Hatta:2013qj} based on the BMS equation \cite{Banfi:2002hw}. We also study isolated photon production and compute the logarithms of $\epsilon_\gamma$, the energy fraction inside the isolation cone. 

The remainder of this paper is organized as follows. In Section \ref{fact} we review the factorization theorem for jet cross sections with gaps or isolation cones. In Section \ref{Rgrun} we will show that RG evolution of the associated Wilson coefficients is equivalent to a parton shower, and we give the necessary ingredients for LL resummation. In Section \ref{pheno} we will apply the shower code to obtain some phenomenological predictions, namely gap fraction of dijet production and isolation cone cross section. We summarize our results and provide some further discussion in Section \ref{conclusion}. 

\section{Factorization for jet cross sections}\label{fact}

Consider an exclusive $k$-jet cross section at a lepton collider with center-of-mass energy $Q$ in which radiation is vetoed in an angular region $\Omega_{\rm out}$ outside the jets. If the veto has an associated energy scale $Q_0$, this process fullfils a factorization formula of the form \cite{Becher:2015hka,Becher:2016mmh} 
\begin{align}\label{sigbarefinal}
d\sigma(Q,Q_0) &=  \sum_{m=k}^\infty \big\langle \bm{\mathcal{H}}_m(\{\underline{n}\},Q,\mu) \otimes \bm{\mathcal{S}}_m(\{\underline{n}\},Q_0,\mu) \big\rangle \,.
\end{align}
The factorization theorem is the leading term in an expansion of the cross section in $\beta= Q_0/Q$. Since the soft radiation is sensitive to the directions  $\{\underline{n}\}=\{n_1,\dots, n_m\}$ and the color charges of the hard partons, both the soft and hard functions depend on these quantities. The symbol $\otimes$ indicates an integral over these directions and $\langle\, \dots\, \rangle$ denotes the color trace, which is taken after combining the two functions. In \eqref{sigbarefinal} we indicate the dependence of the cross section on $Q$ and $Q_0$ explicitly, but it depends on the momenta of the individual jets. The cross section thus involves several individual hard energy scales, but we assume that all of them are of order $Q$ and do not indicate them explicitly. Below, we will compute cross sections as a function of the rapidities and the average transverse momentum of the jets.

The formula \eqref{sigbarefinal} covers a variety of situations. The most common is exclusive jet cross sections, with a veto on additional radiation outside the jets. For low values of the veto scale $Q_0$, the outside region is also called the ``gap'' between the jets. The name ``gap'' refers to studies of forward dijet processes without any hadrons outside the jets \cite{Abe:1994de, Abachi:1995gz,Abe:1997ie}, which is of course problematic in a perturbative context \cite{Oderda:1998en}. For our work, we are interested in values of $Q_0$ in the perturbative domain. Note that the radiation inside the gap is outside the jets; however, throughout our paper ``inside'' will always refer to the region of large energy. A second set of observables obeying \eqref{sigbarefinal} are isolation cone cross sections for small values of the energy inside the cone, which are relevant e.g.\ for photon production. In the above notation $\Omega_{\rm out}$ then refers to the angular region of the isolation cone and  $Q_0$ to the hadronic energy inside it.

The ingredients of the formula \eqref{sigbarefinal} develop large collinear logarithms as the jets become narrow. We have analyzed this situation in \cite{Becher:2015hka,Becher:2016mmh}  and have shown that the hard and soft functions factorize further in this limit. This additional factorization allows for the resummation of the associated logarithms using RG evolution. Concerning the non-global structure this is a purely technical complication, and for simplicity's sake, we will not resum logarithms of the jet radius in the present paper. Such logarithms are of course of interest and were studied in a number of recent papers, both for exclusive and inclusive cross sections, see \cite{Dasgupta:2014yra, Dasgupta:2016bnd,Kolodrubetz:2016dzb,Dai:2016hzf,Kang:2016mcy, Kang:2016ioz,Kang:2017mda}. 

The second, more important limitation of the formula \eqref{sigbarefinal} is that it was derived for $e^+ e^-$ collisions. Naively, one would guess that one simply will need to include a convolution with parton distribution functions (PDFs) to account for  incoming partons and generalize  \eqref{sigbarefinal} to hadron colliders. However, the work of \cite{Forshaw:2006fk,Forshaw:2008cq} has shown that beyond the large-$N_c$ limit, the factorization properties become more complicated. The anomalous dimension which governs the hard function evolution involves Glauber (or Coulomb) phases which no longer cancel in the hadron collider case. This leads to double logarithms at higher orders which cannot be absorbed into PDFs. It will be interesting to analyze the low-energy theory in the presence of these ``super-leading'' logarithms. In the present work we will remain in the large-$N_c$ limit where these complications are absent.

The factorization theorem \eqref{sigbarefinal}  is based on the factorization of soft radiation from a hard amplitude with $m$ partons, which takes the form
\begin{equation}\label{eq:WilsonSoft}
 \bm{S}_1(n_1) \, \bm{S}_2(n_2) \,  \dots\,  {\bm S}_m(n_m)\,|\mathcal{M}_m(\{\underline{p}\})\rangle \,,
\end{equation}
where $ \bm{S}_i(n_i)$ is a Wilson line along the direction of particle $i$ in the appropriate color representation. The soft functions are given by the matrix element squared of emissions from these Wilson lines
\begin{equation}
\bm{\mathcal{S}}_m(\{\underline{n}\},Q_0,\mu) = \int\limits_{X_s}\hspace{-0.55cm}\sum \,\langle  0 |\, \bm{S}_1^\dagger(n_1) \,  \dots\,  {\bm S}_m^\dagger(n_m)\,  |X_s \rangle\langle  X_s | \,\bm{S}_1(n_1) \,  \dots\,  {\bm S}_m(n_m) \, |0 \rangle \, \theta( Q_0 - E_{\rm \, out}) \,,\label{eq:Sn}
\end{equation}
where the states $X_s$ contain an arbitrary number of soft partons. The soft functions depend on the energy $Q_0$ of the radiation and implicitly also on the shape of the region $\Omega_{\rm out}$ in which the energy is measured. The Wilson-line matrix elements have ultraviolet divergences which can be renormalized away and this induces a dependence on the renormalization scale $\mu$. 

The hard functions are given by the square of the hard-scattering amplitudes, together with the phase-space constraints ${\Theta }_{\rm in}\!\left(\left\{\underline{p}\right\}\right)$ which restrict the $m$ hard partons to the inside of the jets, 
\begin{align}\label{eq:Hm}
\bm{\mathcal{H}}_m(\{\underline{n}\},Q,\mu) =\frac{1}{2Q^2} \sum_{\rm spins}
\prod_{i=1}^m & \int \! \frac{dE_i \,E_i^{d-3} }{(2\pi)^{d-2}} \, |\mathcal{M}_m(\{\underline{p}\}) \rangle \langle  \mathcal{M}_m(\{\underline{p}\}) |\nonumber \\
&\times (2\pi)^d \,\delta\Big(Q - \sum_{i=1}^m E_i\Big) \,\delta^{(d-1)}(\vec{p}_{\rm tot})\,{\Theta }_{\rm in}\!\left(\left\{\underline{p}\right\}\right) \,.
\end{align}
For cone jets the phase-space constraint ${\Theta }_{\rm in}\!\left(\left\{\underline{p}\right\}\right)$ is defined by cones around the hard partons. For recombination algorithms, on the other hand, the jet clustering constraints can be quite complicated in general and can spoil factorization. However, they simplify in our setup which considers the limit of hard partons together with (infinitely) soft radiation. This situation was considered in \cite{Cacciari:2008gp} where it was shown that for anti-$k_T$ jets, the jet boundary becomes cone-like so that the theorem \eqref{sigbarefinal} also applies to this case.

Since the cross section must be independent of the scale $\mu$, the scale dependence among the hard and soft functions must cancel. The one for the hard function is driven by the RG equation 
\begin{align}\label{eq:hrdRG}
\frac{d}{d\ln\mu}\,\bm{\mathcal{H}}_m(\{\underline{n} \},Q,\mu)  &= - \sum_{l =k}^{m}  \bm{\mathcal{H}}_l(\{\underline{n} \},Q,\mu) \, \bm{\Gamma}^H_{lm}(\{\underline{n}\},Q ,\mu) \, .
\end{align}
This evolution equation is formally solved by the path ordered exponential
\begin{equation}\label{eq:US}
   \bm{U}(\{\underline{n}\},\mu_s,\mu_h) 
   = {\bf P} \exp\left[\, \int_{\mu_s}^{\mu_h} \frac{d\mu}{\mu}\,
    \bm{\Gamma}^H(\{\underline{n}\},\mu) \right]\,,
\end{equation}
and the resummed cross section is then
\begin{equation}
   d\sigma(Q,Q_0) 
   = \sum_{l=k,\,m\geq l}^\infty  \big\langle \bm{\mathcal{H}}_l(\{\underline{n}\},Q,\mu_h) 
    \otimes \bm{U}_{lm}(\{\underline{n}\},\mu_s,\mu_h)\,\hat{\otimes}\, 
    \bm{\mathcal{S}}_m(\{\underline{n}\},Q_0,\mu_s)\big\rangle \,.
\end{equation}
The condition $m\geq l$ arises because the anomalous dimension matrix is zero below the diagonal, see below. The hat in $\hat{\otimes}$ indicates that one has to integrate over the angles of the $(m-l)$ additional unresolved emissions. For the choice $\mu_h\sim Q$ and $\mu_s\sim Q_0$, the hard and soft functions are free of large logarithms and can be expanded in the respective coupling  constants $\alpha_s(\mu_h)$ and $\alpha_s(\mu_s)$. At leading logarithmic accuracy, we only need these functions at leading power in $\alpha_s$. The soft functions then become trivial $\bm{\mathcal{S}}_m=\bm{1}$ and all higher-multiplicity hard functions are suppressed, $\bm{\mathcal{H}}_m \sim \alpha_s^{m-k}\, \bm{\mathcal{H}}_k$. The cross section thus simplifies to
\begin{equation}\label{eq:LLresum}
d\sigma^{\rm LL}(Q,Q_0) = \sum_{m=k}^\infty \big\langle  \bm{\mathcal{H}}_k(\{\underline{n} \},Q,\mu_h)\, \otimes \,  \bm{U}_{km}(\{\underline{n}\},\mu_s,\mu_h) \,\hat{\otimes}\, \bm{1} \big\rangle \,,
\end{equation} 
where the evolution factor can be evaluated with the leading-order expression for the anomalous dimension $\bm{\Gamma}^H$. We note that the Born-level cross section is given by 
\begin{equation}
d\sigma_{\rm 0}(Q,Q_0) = \big\langle  \bm{\mathcal{H}}_k(\{\underline{n}\},Q,\mu_h) \big\rangle \,.
\end{equation} 
This demonstrates that the starting point of the evolution is the tree-level cross section, as we have indicated earlier. The additional piece of information needed is the color structure since the evolution changes the colors. The paper \cite{Farhi:2015jca} has modified the {\sc MadGraph} code in such a way that it provides the full color information. We will focus on the large-$N_c$ limit below and use the color information which {\sc MadGraph} provides for showering its tree-level events. We will come back to this point later.

\begin{figure}[t]
 \centering
 \includegraphics[width=0.6\textwidth]{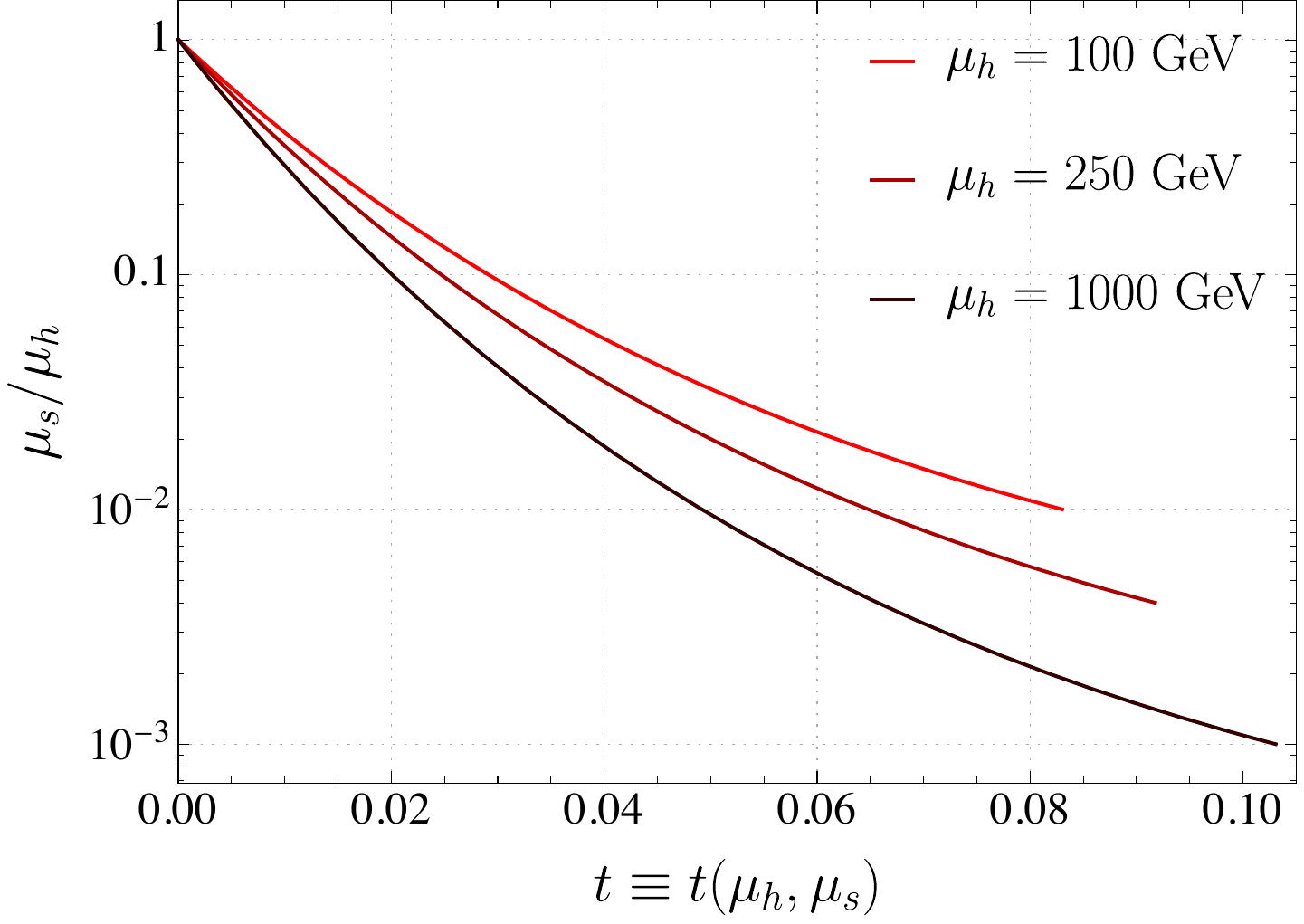}
  \caption{The relation between shower time $t$,  hard scale $\mu_h$ and soft scale $\mu_s$. We stop the lines in the plot when $\mu_s$ reaches $1\,{\rm GeV}$.
 \label{fig:showeringtime}}
 \end{figure}

It is convenient to rewrite the exponent of the evolution matrix \eqref{eq:US}  at leading order in RG-improved perturbation theory in the form
\begin{equation}
\int_{\mu_s}^{\mu_h} \frac{d\mu}{\mu}\, \bm{\Gamma}^H_{nm}= \int_{\alpha(\mu_s)}^{\alpha(\mu_h)} \frac{d\alpha}{\beta(\alpha)}\, \frac{\alpha}{4\pi}\,\bm{\Gamma}_{nm}^{(1)} =\frac{1}{2\beta_0}\ln\frac{\alpha(\mu_s)}{\alpha(\mu_h)}\,\bm{\Gamma}_{nm}^{(1)}  \,.
\end{equation}
Using the one-loop anomalous-dimension matrix $\bm{\Gamma}_{nm}^{(1)}$ yields leading logarithmic accuracy in the evolution. The prefactor
\begin{equation}\label{eq:runt}
t = \frac{1}{2\beta_0}\ln\frac{\alpha(\mu_s)}{\alpha(\mu_h)} =\frac{\alpha_s}{4\pi} \ln\frac{\mu_h}{\mu_s} +\mathcal{O}(\alpha_s^2)
\end{equation}
is the ``evolution time'', which we will call shower time in the context of the parton shower. We start the evolution at $t=0$ and then evolve to larger times, which correspond to lower scales. Since we will sometimes plot quantities as a function of the shower time $t$, we show the relation between $t$ and the ratio of the low scale $\mu_s$ to the high scale $\mu_h$ for different hard-scattering scales $\mu_h$ in Figure \ref{fig:showeringtime}. The plot makes it clear that the relevant region for perturbative calculations is $t \lesssim 0.1$, even after resummation.

\section{RG evolution as a parton shower}\label{Rgrun}

To obtain a MC implementation of the leading-logarithmic evolution we make use of the explicit form of the one-loop anomalous dimension \cite{Becher:2016mmh}, which for $k$-jet production has the form
\begin{equation}\label{eq:gammaOne}
\bm{\Gamma}^{(1)} =  \left(
\begin{array}{ccccc}
   \, \bm{V}_{k} &   \bm{R}_{k} &  0 & 0 & \hdots \\
 0 & \bm{V}_{k+1} & \bm{R}_{k+1}  & 0 & \hdots \\
0 &0  &  \bm{V}_{k+2} &  \bm{R}_{k+2} &   \hdots \\
 0& 0& 0 &  \bm{V}_{k+3} & \hdots
   \\
 \vdots & \vdots & \vdots & \vdots &
   \ddots \\
\end{array}
\right).
\end{equation}
The one-loop anomalous dimensions are given by 
\begin{align}\label{eq:oneLoopRG}
 \bm{V}_m  &= 2\,\sum_{(ij)}\,(\bm{T}_{i,L}\cdot  \bm{T}_{j,L}+\bm{T}_{i,R}\cdot  \bm{T}_{j,R})  \int \frac{d\Omega(n_l)}{4\pi}\, W_{ij}^l 
 \nonumber\\
  &\hspace{2cm} - 2\, i \pi \,\sum_{(ij)} \left(\bm{T}_{i,L}\cdot  \bm{T}_{j,L} - \bm{T}_{i,R}\cdot  \bm{T}_{j,R}\right) \Pi_{ij}
 , \\
 \bm{R}_m & = -4\,\sum_{(ij)}\,\bm{T}_{i,L}\cdot\bm{T}_{j,R}  \,W_{ij}^{m+1}\,  \Theta_{\rm in}(n_{m+1})\,. \nonumber
\end{align}
In \cite{Becher:2016mmh}, they were derived by considering soft limits of the amplitudes. The relevant product of soft currents leads to a dipole structure for the angular dependence given by the factor
\begin{equation}
W_{ij}^l = \frac{n_i\cdot n_j}{n_i\cdot n_l \, n_j\cdot n_l}\,.
\end{equation}
Before discussing the evolution, let us explain how the anomalous dimension acts on the functions $\bm{\mathcal{H}}_m$ defined in \eqref{eq:Hm}. These functions contain both amplitudes $|\mathcal{M}_m(\{\underline{p}\}) \rangle$ and their conjugate. The color matrices $\bm{T}_{i,L}$ acts on the $i$-th parton in the amplitude while $\bm{T}_{j,R}$ multiplies the conjugate, for example
\begin{equation}
(\bm{T}_{1,L}\cdot  \bm{T}_{2,L} + \bm{T}_{3,R}\cdot \bm{T}_{4,R})\, \bm{\mathcal{H}}_m = \bm{T}_{1}\cdot  \bm{T}_{2}\, \bm{\mathcal{H}}_m + \, \bm{\mathcal{H}}_m\, \bm{T}_{3}\cdot \bm{T}_{4} \,.
\end{equation}
and $\bm{T}_{i,L}\cdot  \bm{T}_{j,L}=\sum_a \bm{T}_{i,L}^a\cdot  \bm{T}_{j,L}^a$. This is the usual color-space notation \cite{Catani:1996jh,Catani:1996vz}. While we do not indicate this notationally, the color matrices in the real-emission operator  $\bm{R}_m$ are different. They take an amplitude with $m$ partons and associated color indices and map it into an amplitude with $m+1$ partons. Explicitly, we have
\begin{equation}
\bm{T}_{i,L}\cdot\bm{T}_{j,R} \,\bm{\mathcal{H}}_m = \bm{T}_{i}^a\, \bm{\mathcal{H}}_m \,\bm{T}_{j}^a\,.
\end{equation}
and the index $a$ is the color of the emitted gluon. Note that there is no sum over the color $a$. The color sum will only be taken at the end after multiplying with the soft function. We nevertheless use the scalar product notation $\bm{T}_{i,L}\cdot\bm{T}_{j,R}$ since it allows us to suppress the color indices, which is one of the advantages of the color-space formalism. However, when applying the real emission operator $\bm{R}_m$ one needs to keep in mind that one changes into new color space and that subsequent applications of color matrices can act on the new color index.

We have explicitly indicated the imaginary part of the virtual diagrams in the anomalous dimension \eqref{eq:oneLoopRG}. The corresponding Glauber phase arises from cutting the two lines between which the virtual gluon is exchanged and arises when $i$ and $j$ are both incoming or outgoing, and the factor $\Pi_{ij}$ is defined to be $1$ in this case and $0$ otherwise. For $e^+e^-$ collisions, this part immediately vanishes due to color conservation $\sum_i  \bm{T}_{i}=0$ but it is present in hadronic collisions and induces the super-leading logarithms discovered in \cite{Forshaw:2006fk,Forshaw:2008cq}.

Let us now discuss the solution of the RG at leading logarithmic accuracy. Using the simple structure of the anomalous dimension matrix \eqref{eq:gammaOne} and changing variables from $\mu$ to $t$, the RG equation \eqref{eq:hrdRG} reads 
\begin{align}\label{eq:diffhrd}
\frac{d}{dt}\,\bm{\mathcal{H}}_m(t)  &=   \bm{\mathcal{H}}_m(t) \,  \bm{V}_m +   \bm{\mathcal{H}}_{m-1}(t) \,  \bm{R}_{m-1} \, ,
\end{align}
where we have suppressed the dependence on the other variables. The solution of the homogenous part of the equation is simply an exponential and we can thus rewrite \eqref{eq:diffhrd} as
\begin{equation}\label{eq:MCstep}
\bm{\mathcal{H}}_m(t) = \bm{\mathcal{H}}_m(t_0) \,e^{(t-t_0) \bm{V}_m}
+ \int_{t_0}^{t} dt' \,\bm{\mathcal{H}}_{m-1}(t') \, \bm{R}_{m-1}\, e^{(t-t')  \bm{V}_m}\,.
\end{equation}
This is the form in which parton-shower equations are usually presented: we evolve from $t_0$ to time $t$ either without an emission (the first part), or by adding an additional emission to a lower-leg amplitude. In this context $e^{(t-t') \bm{V}_m}$ is usually called the Sudakov factor, but since our problem is single logarithmic, this nomenclature does not quite fit. To map to expression \eqref{eq:LLresum}, we note that
\begin{align}
\bm{\mathcal{H}}_m(t) \equiv \bm{\mathcal{H}}_k(\{\underline{n}\},Q,\mu_h)\, \bm{U}_{km}(\{\underline{n}\},\mu_s,\mu_h)\,,
\end{align}
and that the initial condition is $\bm{\mathcal{H}}_m(0)=0$ for all $m>k$. To solve the equation for a process with $k$ jets, one starts with $m=k$ and then uses \eqref{eq:MCstep} iteratively to generate all higher functions
\begin{align}\label{eq:iterRG}
\bm{\mathcal{H}}_k(t) &= \bm{\mathcal{H}}_k(0) \,e^{t \bm{V}_k} \,,\nonumber\\
\bm{\mathcal{H}}_{k+1}(t) &= \int_{0}^{t} dt' \,\bm{\mathcal{H}}_{k}(t') \, \bm{R}_{k}\, e^{(t-t')  \bm{V}_{k+1}}\,, \\
\bm{\mathcal{H}}_{k+2}(t) &= \int_{0}^{t} dt' \,\bm{\mathcal{H}}_{k+1}(t') \, \bm{R}_{k+1}\, e^{(t-t')  \bm{V}_{k+2}}\,, \nonumber \\
\bm{\mathcal{H}}_{k+3}(t) &= \dots \,. \nonumber 
\end{align}
To get the resummed result, one evolves to the appropriate value of $t$, which is set by the scales $\mu_h$ and $\mu_s$ in \eqref{eq:runt}. The leading-logarithmic cross section is obtained from the sum
\begin{align}\label{eq:sigmaLL}
d\sigma_{\rm LL}(Q,Q_0) &= \sum_{m=k}^\infty \big\langle  \bm{\mathcal{H}}_m(t) \,\hat{\otimes}\, \bm{1} \big\rangle\nonumber \\
&= \big\langle \bm{\mathcal{H}}_k(t) + \int \frac{d\Omega_1}{4\pi} \bm{\mathcal{H}}_{k+1}(t) +\int \frac{d\Omega_1}{4\pi}\int \frac{d\Omega_2}{4\pi}  \bm{\mathcal{H}}_{k+2}(t) + \dots \big\rangle \,,
\end{align}
where we have explicitly written out the angular integrations over the additional emissions generated by the shower. 

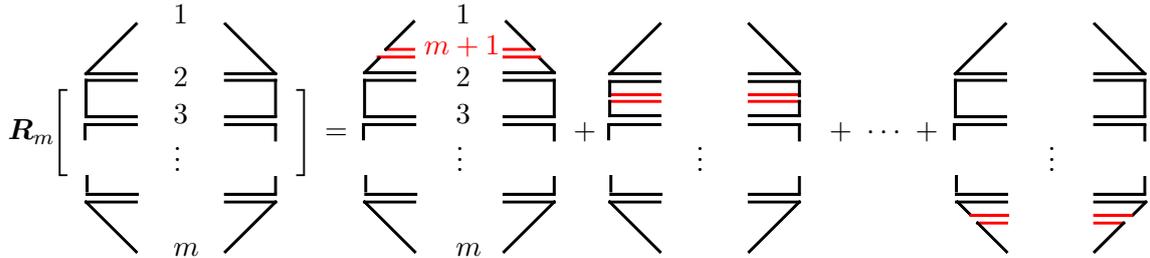
\begin{figure}[t!]
\centering
\hspace{1cm}\begin{overpic}[scale=0.85]{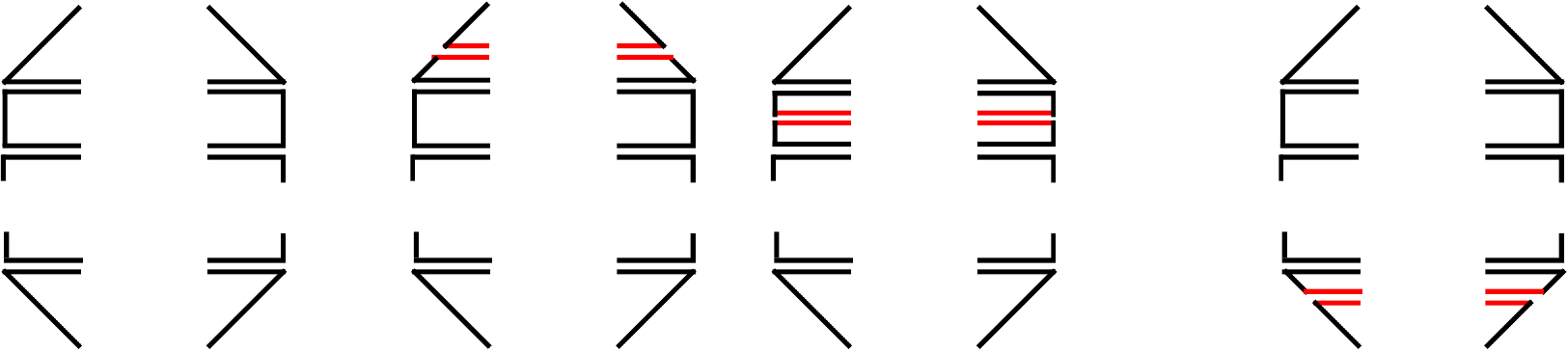}
\put(-7,11){$ \bm{R}_m$}
\put(-3,11){$\Bigg[$}
\put(20,11){$\Bigg]\hspace{0.05cm}=$}
\put(46,11){$+$}
\put(70,11){$+\,\,\cdots\,\,+$}
\put(8.5,22){$1$}
\put(8.5,0){$m$}
\put(8.5,16){$2$}
\put(8.5,12.5){$3$}
\put(8.5,4.5){$$}
\put(32,19){$\color{red}m+1$}
\put(35,22){$1$}
\put(35,0){$m$}
\put(35,16){$2$}
\put(35,12.5){$3$}
\put(35,4.5){$$}
\put(8.5,8){$\vdots$}
\put(35,8){$\vdots$}
\put(57.5,8){$\vdots$}
\put(90.5,8){$\vdots$}
\end{overpic}
\caption{The action of the operator $\bm{R}_m$ on an amplitude with $m$ legs in the large-$N_c$ limit. The double and single lines represent gluons and quarks, respectively. \label{fig:BMS_dia}}
\end{figure}

To perform the integrations over the intermediate times and the angles of the emissions, one has to resort to MC methods. Implementing the above equations is difficult because the hard functions and anomalous dimension are matrices in the color space of the involved partons and the dimension of this space rapidly grows for higher particle multiplicities. For this reason a full implementation of color into a parton shower has so far not been achieved, but there are methods to systematically expand around the large-$N_c$ limit \cite{Platzer:2012np, Platzer:2013fha,Martinez:2018ffw}. Here, we will work in the strict large-$N_c$ limit and use the trace basis for the color structure, so that emissions only arise between neighbouring legs
\begin{equation}\label{eq:largeNc}
\bm{T}_{i}\cdot  \bm{T}_{j} \to -\frac{N_c}{2} \delta_{i,j\pm1}\, \bm{1}\,,
\end{equation}
and each loop or real emission simply leads to an additional factor of $N_c$. We have discussed this point in detail in \cite{Becher:2016mmh} and reproduce an illustration from this paper in Figure~\ref{fig:BMS_dia} which shows how the real-emission operator $\bm{R}_{m}$ acts on an amplitude with $m$ legs. The amplitude at large $N_c$ can be viewed as a set of color dipoles and the real emission operator adds a new leg, splitting an existing dipole into two new ones.  Similarly, the virtual correction operator \eqref{eq:oneLoopRG} reduces to a sum of integrals for each dipole involving neighbouring legs
\begin{equation}\label{eq:virtual}
\bm{V}_m = -4N_c\, \bm{1}\, \sum_{i} \int \frac{d\Omega(n_l)}{4\pi}\, W_{i,i+1}^l 
\end{equation} 
in the large-$N_c$ limit. The treatment of color is of course completely standard and exactly what is implemented in all existing parton-shower programs. In our practical implementation, we work with Les Houches Event Files (LHEF) \cite{Alwall:2006yp} obtained by computing the tree-level amplitudes with
{\sc MadGraph5\Q{_}aMC@NLO}. The event files provide the directions of the hard partons in $\bm{\mathcal{H}}_k(t)$ as well as their color connections. We can thus read out all the necessary information to start the shower and to generate $\bm{\mathcal{H}}_m(t)$ for $m>k$.

Individually both $\bm{R}_{m}$ and $\bm{V}_m$ suffer from collinear divergences. These cancel in physical observables, but need to be regularized in our shower since we want to exponentiate the virtual corrections, see \eqref{eq:MCstep}. A simple way to achieve this is to regularize the dipole as
\begin{equation}
W_{ij}^l   \to  W_{ij}^l \, \theta( n_l \cdot n_i - \lambda^2 ) \, \theta(n_l \cdot n_j - \lambda^2)
\end{equation}
in both $\bm{R}_{m}$ and $\bm{V}_m$. The virtual integral \eqref{eq:virtual} with this regulator is analyzed in detail in Appendix \ref{sec:labcone}. To efficiently generate the real emissions, it is advantageous to use the rapidity $\hat{y}$ and the azimuthal angle $\hat{\phi}$ in the center-of-mass frame of the dipole as integration variables, the details can again be found in the Appendix \ref{sec:labcone}. Another way of regularizing the integrals is to impose a cut on the rapidity $\hat{y}$, as was done by \cite{Dasgupta:2001sh}. In Appendix \ref{sec:MCdetail}, we give a detailed description of the MC algorithm and compare the different cutoffs.

\section{Phenomenology of non-global observables}\label{pheno}

In this section we use our simulation code for phenomenological studies and analyze the numerical impact of the resummation for gaps between jets and  isolation cone cross sections for photon production. We will also explain why NGLs for jet-veto cross sections are negligible for the cut parameters used at the LHC.

\subsection{Qualitative discussion\label{sec:quali}}

Before we perform detailed studies, it is useful to start with a qualitative discussion of the size and form of the leading NGLs. For concreteness, let us consider a dijet cross section in $e^+e^-$ with a gap of size $\Delta y$ between the jets, in which radiation above an energy $Q_0$ is vetoed.  This interjet energy flow is the poster child of a non-global observable and was studied for example in \cite{Berger:2001ns, Dasgupta:2002bw, Becher:2016mmh}. 

If the soft radiation would arise entirely from the two Wilson lines associated with the original partons, the leading logarithms would exponentiate as
\begin{align}\label{eq:global}
\frac{\sigma_{\rm GL}^{\rm LL}}{\sigma_0} = \exp\left( - 8  \, C_F \, t \,\Delta y \right),
\end{align}
where the variable $t=\frac{\alpha_s}{4\pi} \ln \frac{Q}{Q_0}$ up to running coupling effects, see \eqref{eq:runt}. For dijet production, these logarithms arise from $\bm{\mathcal{S}}_2$ and are called {\em global} to distinguish them from the complicated pattern from the operators with more Wilson lines. One observes that for these global contributions, each large logarithm is multiplied by the size of the gap $\Delta y$, which is of course expected since one has to recover the inclusive cross section as the gap size becomes zero. In the opposite limit, the prefactor $\Delta y \to \infty$ corresponds to the collinear logarithm which multiplies the soft logarithm present in $t$. The quantity shown in \eqref{eq:global}, the ratio of the cross section with a rapidity gap to the inclusive cross section, is called the gap fraction and corresponds to the fraction of events with radiation in the gap below the veto-scale $Q_0$. 

\begin{figure}[t]
 \centering
 \includegraphics[width=0.47\textwidth]{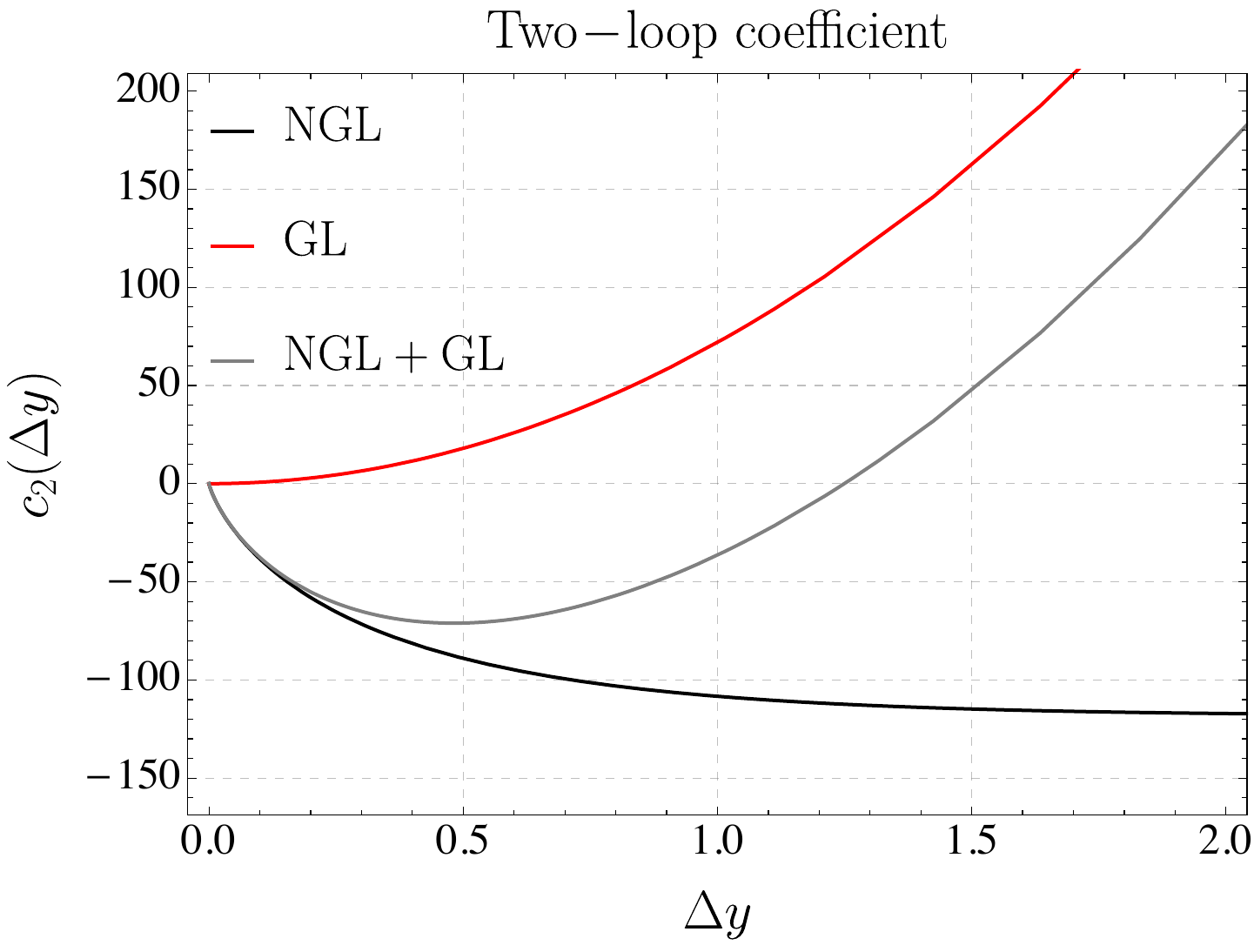}
 \includegraphics[width=0.47\textwidth]{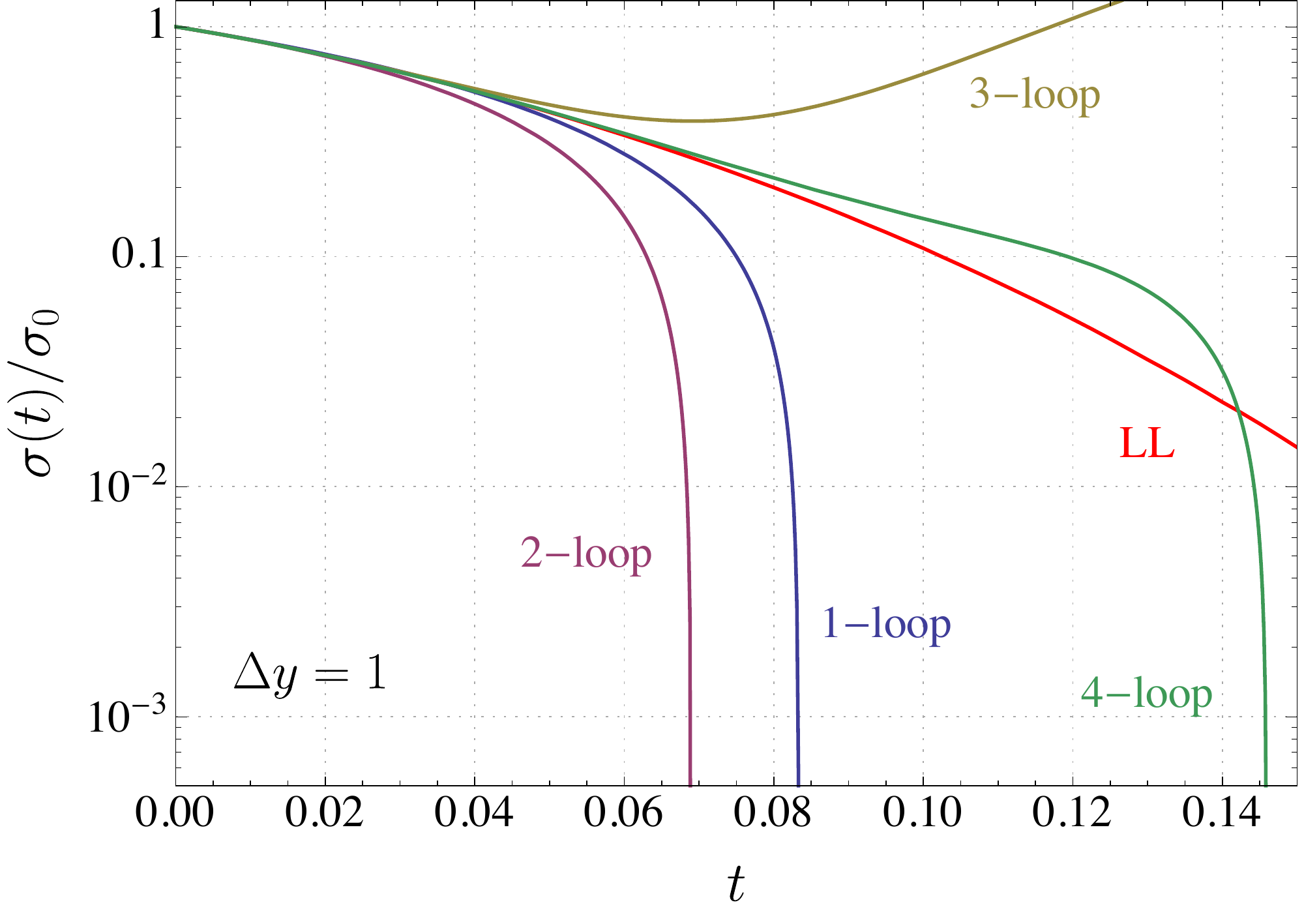}
 \caption{ Left: Two-loop global and non-global coefficients as a function of the gap size $\Delta y$. Right: Comparison of the LL resummation and fixed-order results up to four loops, for $\Delta y=1$.  \label{fig:qualitative_fig}}
 \end{figure}

The leading NGLs to the same observable arise at two-loops and are given by \cite{Dasgupta:2002bw, Becher:2016mmh}
\begin{align}\label{ngl2loop}
\frac{\sigma_{\rm NGL}^{\rm LL}}{\sigma_0} = 4 \, C_F C_A \left[ - \frac{2\pi^2}{3} + 4\, {\rm Li}_2\left( e^{-2\Delta y} \right) \right] t^2.
\end{align}
This contribution arises from a hard gluon emission inside one of the jets, which in turn emits a soft gluon into the gap between the jets. It is encoded in the term $ \bm{\mathcal{H}}_3 \otimes \bm{\mathcal{S}}_3$ in the factorization formula \eqref{sigbarefinal}.

In Figure~\ref{fig:qualitative_fig}, we numerically compare the two-loop global and non-global coefficients as a function of the gap size $\Delta y$, working in the large-$N_c$ limit. When the veto area is small, the gap fraction is dominated by the non-global part, but with increasing veto area the global logarithms become more and more important. Since the two contributions have opposite sign, cancellations between global and non-global contributions can occur at intermediate values of the gap size. To understand this behavior better, it is instructive to expand (\ref{ngl2loop}) in the small $\Delta y$ region
\begin{align}\label{expNGL}
\frac{\sigma_{\rm NGL}^{\rm LL}}{\sigma_0} = 4 \, C_F C_A \Big [  
8 \,\Delta y \big(  \ln(2 \Delta y)  -1 \big)  
- 4 \, \Delta y^2 + \dots \Big ] t^2 \,  .
\end{align}
The expansion \eqref{expNGL} shows that the two-loop non-global logarithmic term is only suppressed by a single power of $\Delta y$, while the global piece involves two powers. The reason for this scaling is that in the non-global piece only one gluon is in the gap of size $\Delta y$, while in the global piece both gluons are. One further observes that in the large-$N_c$ limit the $ \Delta y^2$ part of the non-global piece precisely cancels the global piece. Phenomenologically, the limit of a small gap is for example relevant for isolation cone cross sections, where the veto typically is only applied in a small angular region. Below, we will see an explicit example where the higher-order global and non-global effects cancel for a photon isolation cross section. 

Interestingly, the leading term in \eqref{expNGL} involves a logarithm of $\Delta y$. This contribution corresponds to a collinear enhancement which arises when both the gluon in the gap and the one outside are close to the boundary. These types of collinear logarithms were studied in the recent paper \cite{Hatta:2017fwr} which presented a version of the BMS equation which allows for their all-order resummation. It would be interesting to analyze this in our effective field theory framework. The corresponding effective theory would involve boundary modes to describe the emissions near the gap boundary. The problem is however challenging because the gap fraction is suppressed by a power of $\Delta y$ in the limit  $\Delta y \to 0$.

\subsection{Gaps between jets}

\begin{figure}[t!]
\begin{center}
\begin{overpic}[scale=0.6]{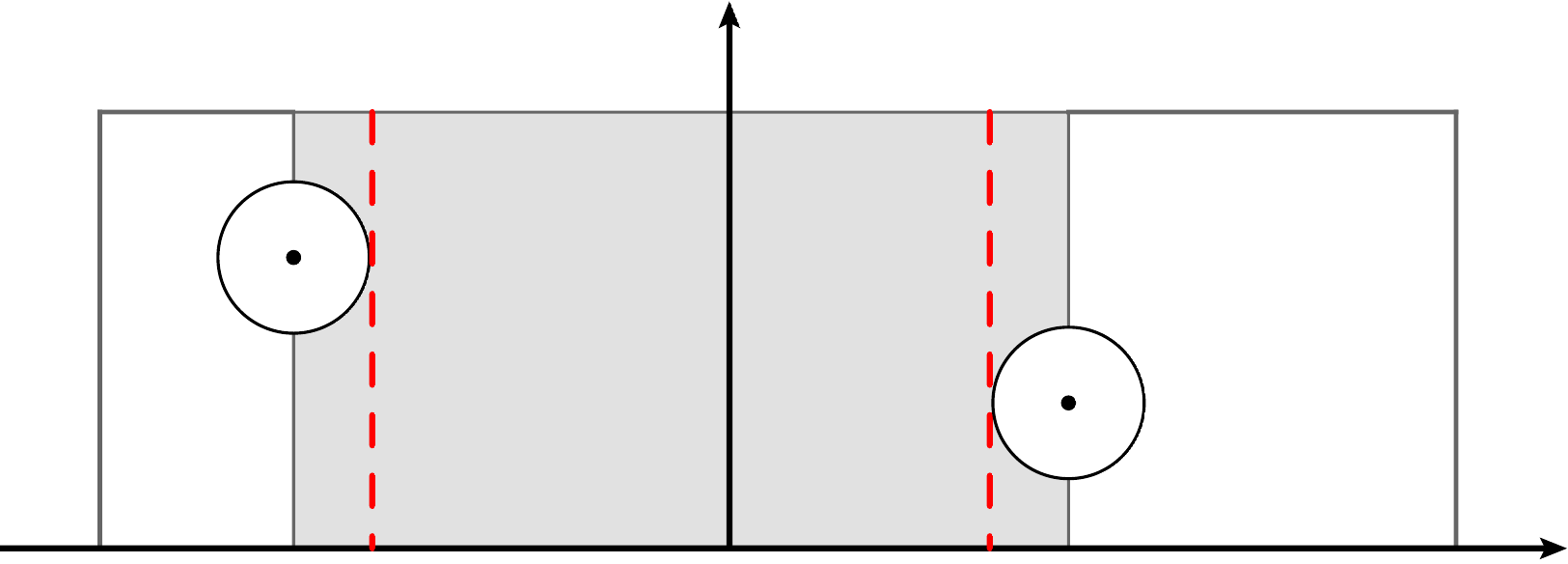}
\put(100,-3){$y$}
\put(50,36){$\phi$}
\end{overpic}
\end{center}
\vspace{-0.3cm}
\caption{Definition of the gap region for a dijet system in the rapidity and azimuthal plane, as used by ATLAS \cite{Aad:2011jz}. If a jet with transverse momentum larger than $Q_0$ is radiated into the gray region, the event is vetoed. The two dashed red lines indicate the boundary of the approximated veto region used in \cite{Hatta:2013qj}.\label{fig:gap_veto}}
\end{figure}

%
%

We now perform the resummation for the gap fraction at the LHC, as measured by the ATLAS experiment \cite{Aad:2011jz,Aad:2014pua}.  The gap fraction is defined as the fraction of dijet events that do not have an additional jet with transverse momentum greater than a given veto scale $Q_0$  in the rapidity interval bounded by the dijet system, and we will study it as a function of $\overline{p}_T$, the average transverse momentum of the two leading jets.
More explicitly, the gap fraction is defined as the ratio of the cross sections with and without veto, 
\begin{align}\label{Rdef}
R( \overline{p}_T,Q_0) = \frac{\sigma_{\rm 2-jet}(\overline{p}_T,Q_0)}{\sigma_{\rm 2-jet}(\overline{p}_T,Q_0=\overline{p}_T)}\,.
\end{align}
Since $\overline{p}_T$ is computed using the two leading jets, the transverse momentum of the jet inside the gap is by definition smaller than  $\overline{p}_T$ so that the denominator in the formula is simply the inclusive two jet cross section. Below, we will compute $R(\overline{p}_T,Q_0)$ for different gap sizes defined by the rapidity difference $\Delta y$ between the two leading jets. The precise geometry of the gap is shown in Figure~\ref{fig:gap_veto}. The jets are reconstructed with the anti-$k_T$ jet algorithm with $R=0.6$ and are required to have rapidity $|y|<4.4$. 

The ATLAS paper \cite{Aad:2011jz} observed that MC predictions are not always consistent with ATLAS data.  For example the NLO predictions matched to PYTHIA \cite{Sjostrand:2006za} and HERWIG \cite{Bahr:2008pv} using POWHEG \cite{Alioli:2010xd} are lower than the experimental data, especially in the region of  large $\overline{p}_T$ and rapidity difference $\Delta y$ between the jets. Specifically, for $210~{\rm GeV}<\overline{p}_T<240~{\rm GeV}$ and $4<\Delta y<5$, POWHEG+HERWIG underestimates the data by about $40\%$, and POWHEG+PYTHIA by about $20\%$. 

For small values of $Q_0$, the gap fraction $R(\overline{p}_T, Q_0)$ involves large logarithms of the form $\alpha_s^n \ln^m\overline{p}_T/Q_0$. It is interesting to perform systematic soft gluon resummations to try to understand the difference between theoretical prediction and experimental data. The resummation of the leading logarithms has been studied in the papers \cite{Banfi:2010pa,DuranDelgado:2011tp,Hatta:2013qj}. In \cite{Banfi:2010pa,DuranDelgado:2011tp} the authors resummed all global logarithms with full colour information and the non-global effects were included by reweighting with a $K$ factor. The most detailed theoretical study so far was \cite{Hatta:2013qj}, which resummed all large logarithms at LL in the large-$N_c$ limit by solving the BMS equation and also compared directly to the experimental measurement.  One limitation of this work is that the veto region was approximated by a rectangle in the rapidity and azimuthal angle plane, see Figure~\ref{fig:gap_veto}. This made it possible to obtain all NGLs by boosting the same solution of the BMS equation. In our computation we will take into account the exact veto region used by ATLAS. Rather than relying on the BMS equation, we will use our parton shower to resum the large logarithms. 


Formula \eqref{sigbarefinal} was derived for leptonic collisions. The factorization formula for dijet production at hadron colliders also includes PDFs $f_a(x, \mu)$ and has the form 
\begin{align}\label{hadronfact}
 \frac{d\sigma(Q_0)}{ d\Delta y \, d\,\overline{p}_T} = & \sum_{a,b\,=\,q, \bar{q}, g}\int d x_1 d x_2 \, f_a(x_1, \mu) f_b(x_2, \mu)  \nno \\
&\hspace{2cm} \times \sum_{m=2}^\infty \big\langle  \bm{\mathcal{H}}_m^{ab}(\{\underline{n}\}, \hat{s}, \overline{p}_T,\mu)\, \otimes\, \bm{\mathcal{W}}_m(\{\underline{n}\},\overline p_T,Q_0,\mu) \big\rangle\,,
\end{align}
where $\hat{s}= x_1 x_2 s $ is the partonic center-of-mass energy. The functions $\bm{\mathcal{W}}_m(\{\underline{n}\},\overline p_T,Q_0,\mu)$ consist of a matrix element of the Wilson lines in the operator $\bm{\mathcal{S}}_{m+2}$ for the incoming and outgoing partons, together with collinear fields of the two incoming ones.  The functions $ \bm{\mathcal{W}}_m$ contain rapidity logarithms due to Glauber gluon exchanges, which induce a dependence on the large scale $\overline p_T$. This dependence has to be present in order to cancel the scale dependence of the super-leading logarithms mentioned in Section~\ref{Rgrun}. These double logarithms of $\mu/\overline p_T$ arise from evolving the hard function and have a scale dependence which cannot cancel against the single-logarithmic scale dependence of the purely soft matrix element and the PDFs. We will discuss the factorization for hadron-collider observables in detail in a forthcoming paper. For the moment, we will concentrate on the leading logarithms in the large-$N_c$ limit, where these complications are absent and the resummed cross section takes the simple form
\begin{align}
\frac{d\sigma(Q_0)}{ d\Delta y \, d\,\overline{p}_T} = \sum_{a,b\,=\,q, \bar{q}, g}\int d x_1 d x_2 f_a(x_1, \mu_f) f_b(x_2, \mu_f) H_2^{ab}(\hat{s},\Delta y, \overline{p}_T,\mu_h) \langle U_{2m}(\mu_s,\mu_h) \hat\otimes 1 \rangle\,.
\end{align}
The hard function $H_2^{ab}$ accounts for the process with two partons in the final state, and all kinematics and color information is encoded in the hard events generated by  {\sc MadGraph}. The tree-level generator computes the exact color dependence of the amplitudes, but to interface with a parton shower such as {\sc PYTHIA}, it randomly assigns a possible large-$N_c$ dipole color structure to each tree-level event. We use this color information to start our shower, which then computes the evolution from $2$ partons in the final state to $m$ partons, as encoded in the matrix elements $U_{2m}$ defined in \eqref{eq:US}. Since we use full tree-level amplitudes, our hard function also contains terms of subleading color. The paper \cite{Farhi:2015jca} has modified {\sc MadGraph} in such a way that the full color information is written into the event file. Using this, one could perform a computation in the strict large-$N_c$ limit.

We choose $\mu_f=\mu_h = \overline{p}_T$ as the central values for the factorization and hard scales, and set the soft scale to be $\mu_s = Q_0$.  A lower value of $\mu_f$ would enhance the gap fraction and bring our results closer to the ATLAS measurements. However, the high value is appropriate since the hard anomalous dimension has two parts, a soft contribution related to non-global logarithms and a collinear part inducing the usual Altarelli-Parisi evolution. In our shower, we only evolve with the soft part of the anomalous dimension and to avoid the necessity for additional collinear evolution we have to evaluate the PDFs at the high scale.

\begin{figure}[t!]
\begin{center}
\begin{tabular}{ccc}
\hspace{-0.5cm}
\includegraphics[width=0.46\textwidth]{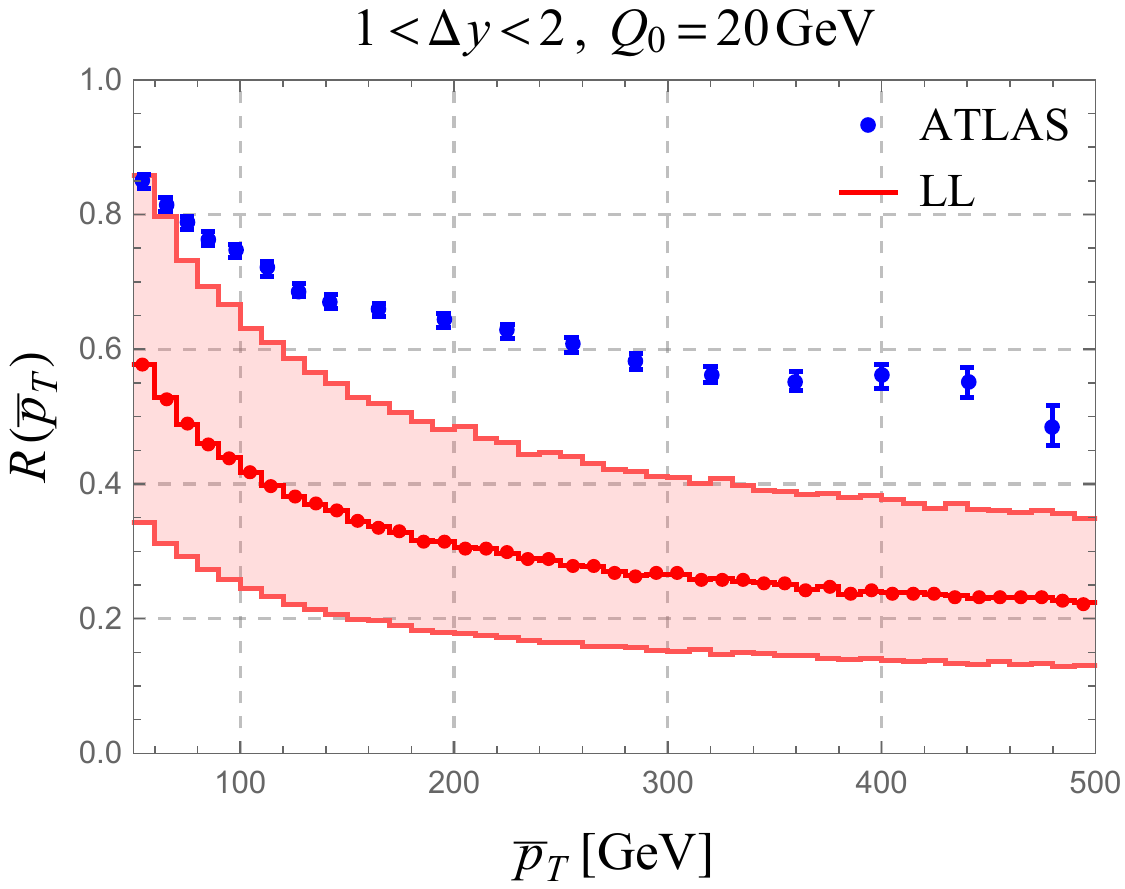} &&  \includegraphics[width=0.46\textwidth]{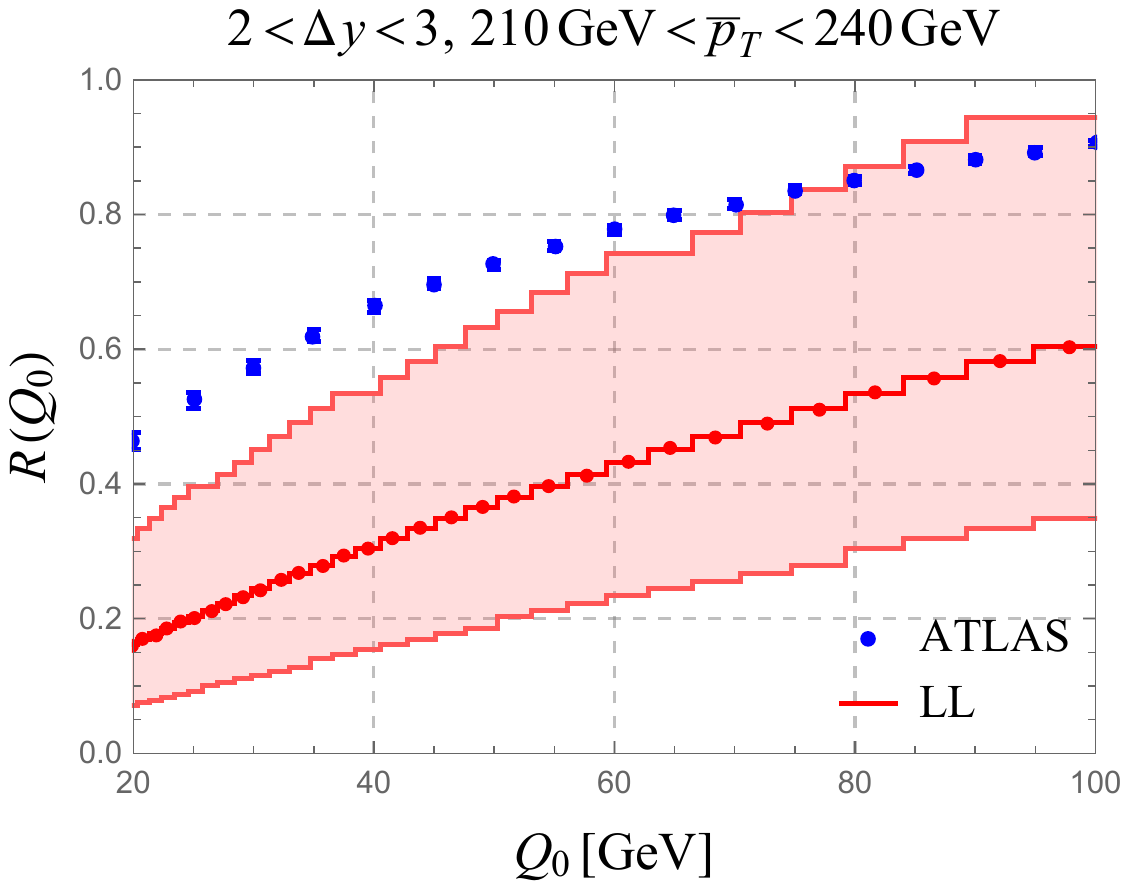}
\end{tabular}
\end{center}
\vspace{-0.3cm}
\caption{The gap fraction as a function of the jet transverse momentum $\overline{p}_T$ (left plot) and the gap energy $Q_0$ (right plot). The red line shows the LL result for the gap fraction; the error band is obtained from scale variation. The {\sc ATLAS} data is plotted in blue.\label{fig:gapLL}}
\end{figure}

\begin{figure}[t!]
        \begin{center}
               \begin{tabular}{c}
                       \hspace{-0.5cm}
                       \includegraphics[width=0.47\textwidth]{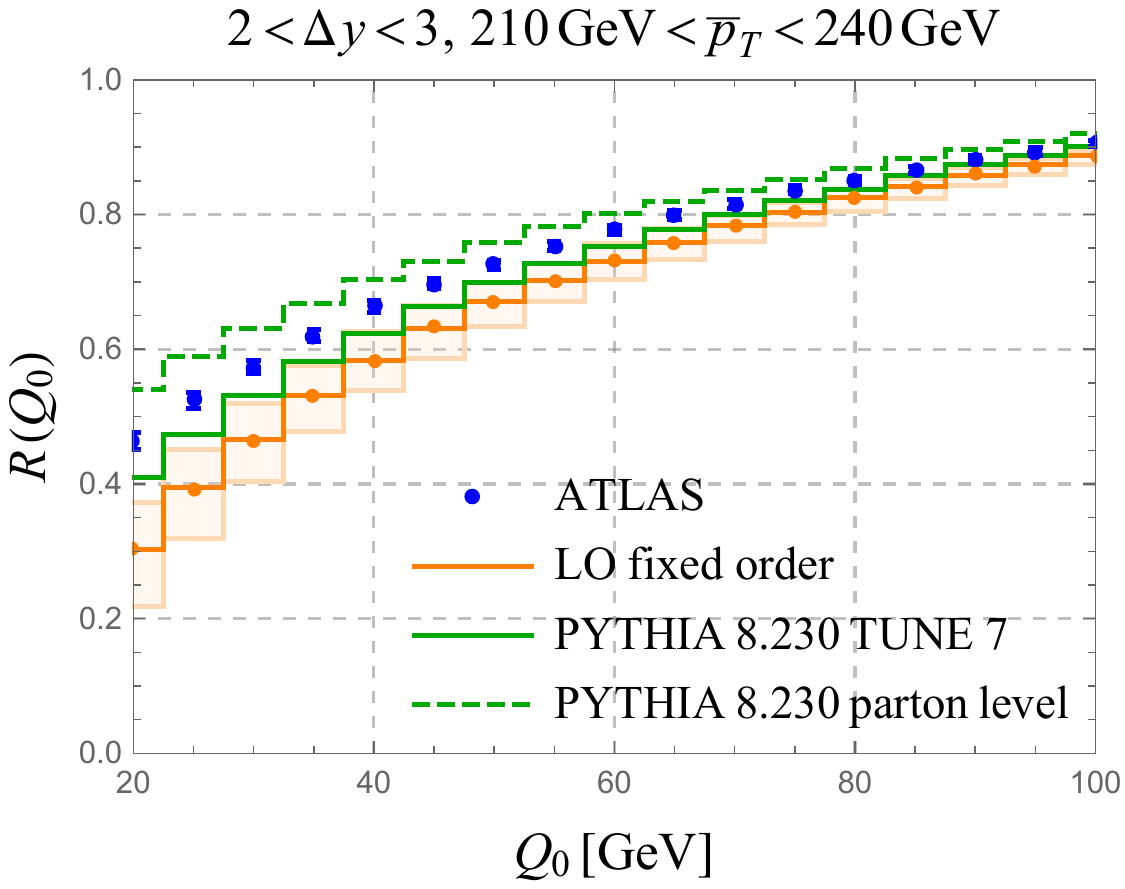} 
                                      \end{tabular}
        \end{center}
        \vspace{-0.3cm}
        \caption{The gap fraction for different gap energies $Q_0$ as measured by {\sc ATLAS} (blue) compared to the fixed-order result at LO (orange) and {\sc PYTHIA} results (solid green: with hadronization using Tune 7, dashed green: partonic result without hadronization and underlying event).\label{fig:gapPythia}}
\end{figure}

In our calculations we use NNPDF23LO  \cite{Ball:2012cx} PDF sets with $\alpha_s(m_Z)=0.130$ and use one-loop running for $\alpha_s$.  In Figure \ref{fig:gapLL} we show the resummed gap fraction in comparison with the ATLAS measurements \cite{Aad:2011jz}. In the left plot, we keep $Q_0=20\,{\rm GeV}$ fixed and vary the transverse momentum $\overline{p}_T$ of the jets, while the right plot shows the gap fraction as a function of $Q_0$ for $210~{\rm GeV}<\overline{p}_T<240~{\rm GeV}$. ATLAS has performed measurements for different rapidity separations between the jets. We want to avoid collinear enhancements and focus on fairly central jets, since we do not resum collinear logarithms for the time being. Specifically, we use $1< \Delta y < 2$ in the left plot and  $2< \Delta y < 3$ in the right one. To estimate the uncertainty of our predictions we vary the scales $\mu_h$ and $\mu_s$ by a factor of two around their default values $\mu_h=\overline{p}_T$ and $\mu_s=Q_0$. The $\mu_s$ variation is larger, except at low $\overline{p}_T$. In the plots we show the envelope of the two variations. We observe that the results are marginally compatible with the experimental measurements within the fairly large uncertainty bands, but it is clear that the theoretical description at LL accuracy is fairly poor. This should be contrasted to the $\mathcal{O}(\alpha_s)$ fixed-order result shown in orange and the result obtained with {\sc PYTHIA} \cite{Sjostrand:2014zea} (solid green line) shown in Figure \ref{fig:gapPythia}. We will call the $\mathcal{O}(\alpha_s)$ prediction leading order (LO), even though strictly speaking  the leading-order gap fraction is $R(\overline{p}_T, Q_0) =1$. Neither the fixed-order result nor {\sc PYTHIA} describe the ATLAS perfectly, but both yield a better description than the LL result. (In their paper ATLAS uses POWHEG matched {\sc PYTHIA}, which agrees with the data well for this rapidity range, but starts deviating at higher rapidities.)

Before speculating about the source of the poor agreement of the LL result with the measurement, it is interesting to compare to \cite{Hatta:2013qj}, which also computed the gap fraction at LL accuracy and compared to the ATLAS data. Superficially, the results presented in this paper show better agreement with data. The reason is two-fold. First of all, the authors not only show the data of the measurement where the gap is defined by the two most energetic jets, but also the experimental results for the case where the gap and $\overline{p}_T$ is defined by the two most forward and most backward jets. This second criterion leads to lower gap fractions, which agree better with the LL resummed result, but -- as the authors of \cite{Hatta:2013qj} readily admit -- is not really appropriate to be compared against the theoretical predictions. Choosing the two most forward and backward jets to define the gap implies a veto on further radiation in the forward and backward direction, which is not imposed in the theoretical computation. Using the highest-$p_T$ jets to define the dijet system, also their gap fractions are below the measurements. They are somewhat higher than our results because \cite{Hatta:2013qj} approximates the gap by a rectangular region in the rapidity and azimuthal angle, see Figure \ref{fig:gap_veto}, so their veto region is smaller than the experimental gap by about one unit of rapidity (the jet radius is $R=0.6$), which increases their gap fraction and brings it closer to data. Adopting their definition of the gap region, we find that our results are consistent with their findings; the remaining small numerical differences can be attributed to the fact that they work in the strict large-$N_c$ limit, while we include the full result for the tree-level amplitudes.

Of course, our computation in the large-$N_c$, leading-logarithmic approximation is rather crude. There are several sources of corrections which could push the results closer to the experimental results. They are (a) higher-logarithmic terms, such as the constant pieces of the one-loop hard and soft functions, (b) power corrections suppressed by $Q_0/\overline{p}_T$, (c) terms of subleading color, or (d) hadronisation and underlying event corrections. Let us rule out the last possibility first. In the experimental measurement, the gap energy $Q_0$ is not defined as the total energy or transverse momentum inside the jet, but as the transverse momentum of the leading jet inside the gap. This definition was chosen to reduce sensitivity to hadronisation and underlying event. Indeed, running {\sc PYTHIA} at the partonic level (dashed green line in Figure \ref{fig:gapPythia}) yields quite similar results to the full simulation (solid green line). We also doubt that subleading-color pieces can explain the difference. Theoretically, the finite-$N_c$ corrections are especially interesting in our case, because at subleading color one encounters double-logarithmic effects, while the problem is only single logarithmic in the large-$N_c$ limit. However, since the double logarithmic effects only arise at $\alpha_s^4$, we do not expect them to be very large. The numerical impact of the super-leading logarithms was estimated to be small in \cite{Forshaw:2009fz}, but one should resum them in order to properly asses their importance. 

\begin{figure}[t!]
        \begin{center}
               \begin{tabular}{ccc}
                       \hspace{-0.5cm}
                  \includegraphics[width=0.45\textwidth]{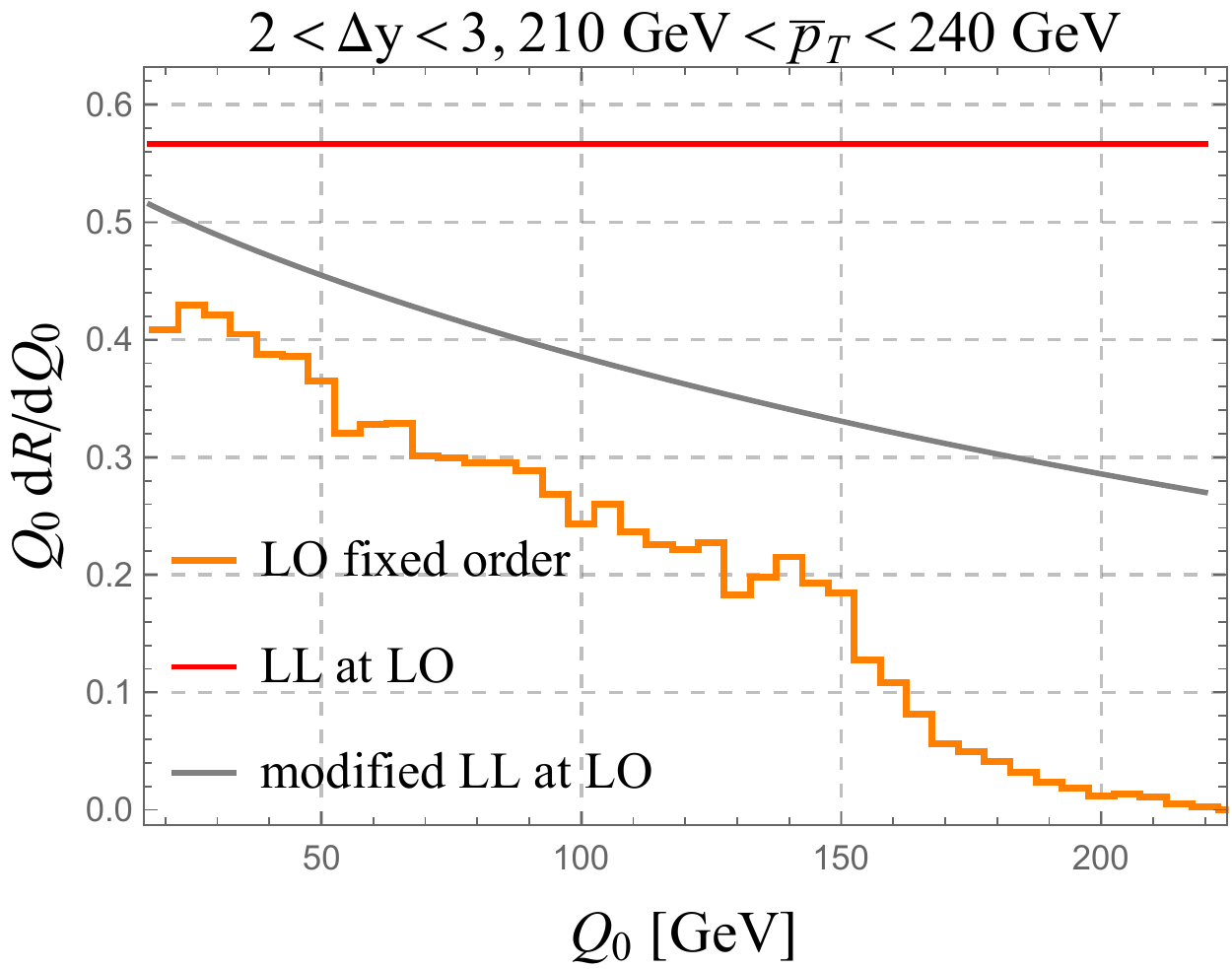}     && \includegraphics[width=0.45\textwidth]{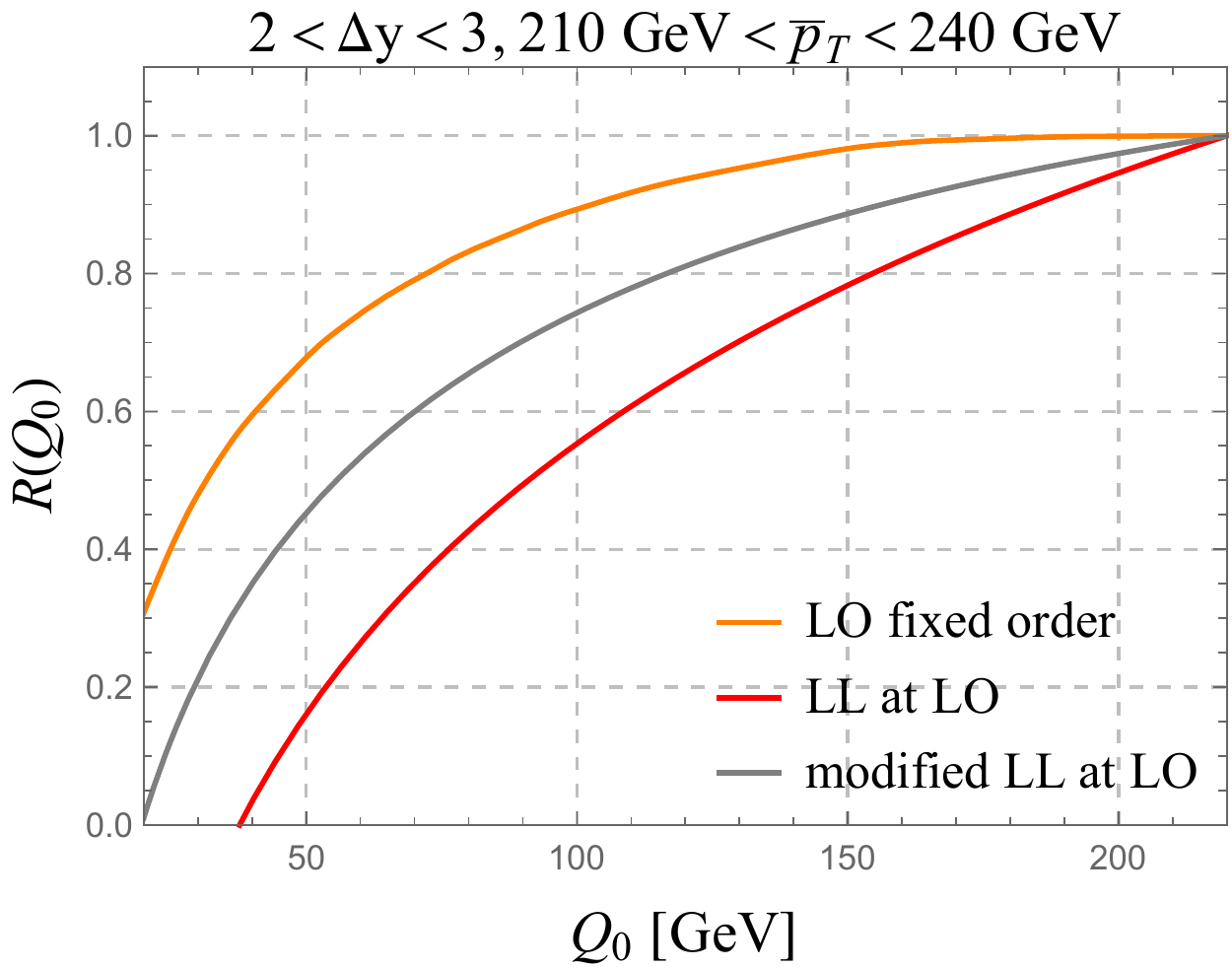} 
               \end{tabular}
        \end{center}
        \vspace{-0.3cm}
        \caption{One emission at LL accuracy, compared to the full LO result. The modified LL shown as a gray line is obtained by implementing momentum conservation for the soft emission.}
        \label{fig:LLtoLO}
\end{figure}

This leaves (a) and (b) as explanations. The scale hierarchy in our computation is not very large $Q_0/\overline{p}_T\gtrsim 1/10$, nevertheless, we expect the power corrections (b) to be moderate. To test their size, we compare in Figure \ref{fig:LLtoLO} the fixed order result at $\mathcal{O}(\alpha_s)$ to the expansion of the LL result to the same accuracy. We compute the LO fixed order result using the relation
\begin{equation}\label{eq:RLO}
R( \overline{p}_T, Q_0)  = 1 - \frac{1}{\sigma^{\rm LO}_{\rm 2-jet}(\overline{p}_T)} \int_{Q_0}^{\overline{p}_T}\!dQ_0'\, \frac{d\sigma^{\rm LO}_{\rm 3-jet}(\overline{p}_T,Q_0')}{dQ_0'}\,.
\end{equation}
At LO, the integrand in \eqref{eq:RLO} is obtained by computing the tree-level three-jet cross section in which the third jet is inside the gap and has transverse momentum $Q_0$. To see the power corrections, it is interesting to take the logarithmic derivative of the gap fraction $R(\overline{p}_T, Q_0)$ with respect to $Q_0$. This removes any constant so that we directly see the difference of the leading-power log term to the full result. As it should be, the full LO result (orange line) approaches the LL coefficient (red line) for small $Q_0$. At the same time the plot shows that the LL derivative is completely off at large $Q_0$, where the derivative of the full LO tends to zero. The fact that $R$ becomes constant at large $Q_0$ implies that power corrections must cancel against the leading-power terms in this region. More generally, the unitarity condition $R(\overline{p}_T, Q_0=\overline{p}_T)=1$ links power corrections (b) and higher-logarithmic terms (a).

One type of power suppressed terms arises from expanding away the soft momenta in the momentum-conservation $\delta$-functions. In our factorization theorem, the momenta in the hard functions at the high scale are conserved, but the soft momenta are neglected. Neglecting the soft momentum $k_s$ enhances the three-jet rate in \eqref{eq:RLO} because the jets can then be produced at the low partonic center-of-mass energy $\hat{s} = (p_{J_1}+ p_{J_1})^2$ instead of the correct value $\hat{s} = (p_{J_1}+ p_{J_1}+k_s)^2$ at which the PDFs are smaller due to the suppression of larger momentum fractions. To gauge the size of this effect, we have used our MC code to compute $dR/dQ_0$ for the first emission with the full $\hat{s}$. Since we know the $k_T=Q_0$ of the emission as well as the direction, we can reconstruct the vector $k$ and the associated $\hat{s}$. In practice, we first boost to the partonic center-of-mass frame, correct $\hat{s}$ and then boost back. Doing so, we obtain the gray line in Figure \ref{fig:LLtoLO}. The modification due to momentum conservation accounts for about half of the difference between LL and the full LO. A similar study was performed in \cite{DuranDelgado:2011tp} who found that they could reproduce the full LO result with good accuracy with a suitable modification of the parton luminosity. However, their modification involved parameters which were chosen by hand. Parton showers such as {\sc PYTHIA} implement momentum conservation, so that these types of kinematic power corrections are accounted for and their effect was also studied in the recent paper \cite{Hoeche:2017jsi}. It is significant, but by itself not large enough to account for the difference we observe. It would be quite interesting to see whether one can modify our shower in such a way that momentum conservation is fulfilled without modifying the leading power terms but we will not pursue this issue further for the moment. 

What can and certainly should be done is to extend the resummation to subleading logarithmic accuracy. This will add the virtual corrections to $\bm{\mathcal{H}}_2^{ab}$ and the function $\bm{\mathcal{H}}_3^{ab}$ at the high scale, together with the $\mathcal{O}(\alpha_s)$ corrections for all the soft functions at the low scale. It will also require the two-loop anomalous dimension in the evolution to lower scales. Computing these corrections and implementing them into a MC is of course a major undertaking. To get a feeling for their size, one can first evaluate the NLL result at $\mathcal{O}(\alpha_s)$. One reason that the higher-log terms are significant is that we have not resummed collinear logarithms for the moment, but with $\Delta y = 3$, these are already of the same order of magnitude as the soft logarithms. Using the results \cite{Becher:2015hka,Becher:2016mmh} this can be done and we plan to implement also the collinear resummation in the future. A related issue is that large rapidity differences lead to forward-scattering kinematics at hadron colliders, which induces its own logarithmic enhancements. A method to resum these terms was put forward in \cite{Andersen:2011hs} and implemented in the {\sc HEJ} code. Recently, the {\sc HEJ} results were merged with  {\sc PYTHIA} \cite{Andersen:2017sht}. This combines both types of resummations and improves the description of the ATLAS data, but to improve our understanding of gap observables, it will be important to perform measurements for kinematical situations in which only a single source of large logarithms is present so that one can separately study the different effects.

\subsection{Isolation cone cross sections and photon production}

\begin{figure}[t!]
\begin{center}
\begin{overpic}[scale=0.5]{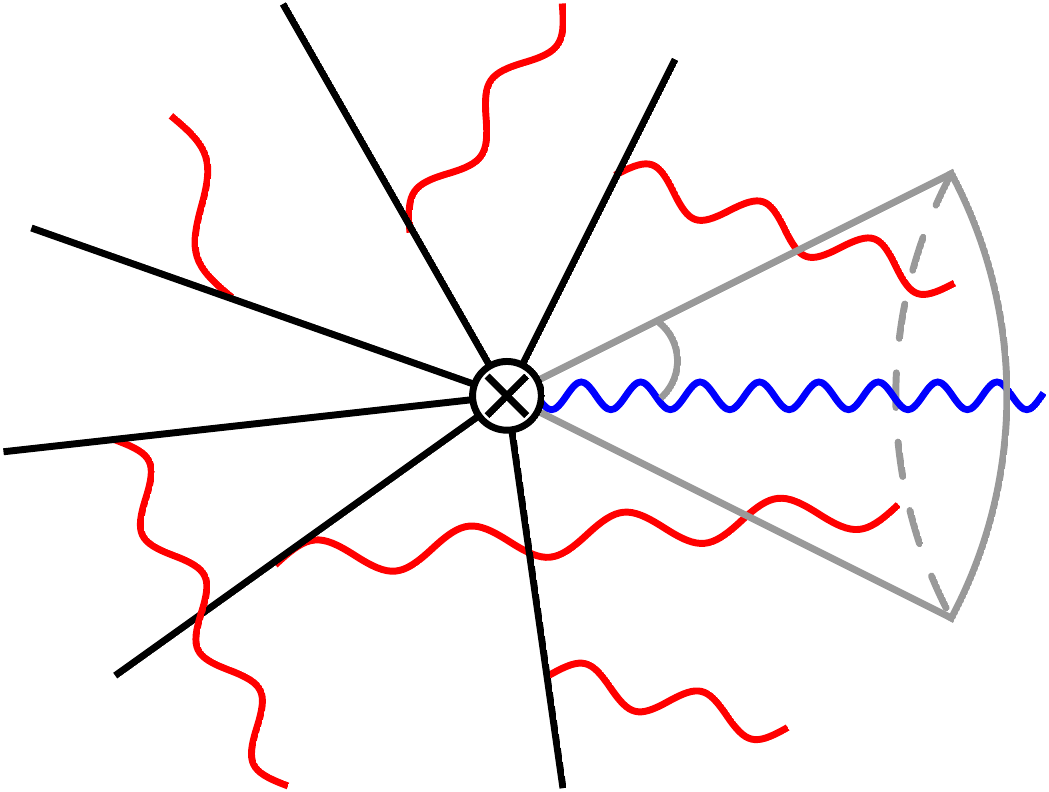}
\put(101,35){$\gamma$}
\put(66,41){$\delta_0$}
\put(101,22){$E_{\rm cone}< E_{\rm iso} = \epsilon_\gamma\,  E_\gamma$}
\end{overpic}
\end{center}
\caption{Pictorial representation of the factorization for isolated photon production. The black lines represent hard partons, while the wavy red lines indicate soft radiation. The energy inside the isolation cone of half-angle $\delta_0$ is restricted to be smaller than $\epsilon_\gamma\,  E_\gamma$.\label{fig:phiso_def}}
\end{figure}

A second important class of non-global observables are cross sections with isolation cones inside which only soft hadronic radiation is allowed. The most important example is photon production, where an isolation cone is needed to separate the direct production of a photon in the underlying hard collision from the photons which arise in hadron decays such as $\pi^0 \to \gamma\gamma$. Imposing that $E_{\rm iso}$, the hadronic energy inside the cone with half-opening angle $\delta_0$, is much smaller than the photon energy $E_\gamma$ suppresses energetic photons originating from decays of boosted hadrons. Similar cuts are also used to isolate leptons, for example in SUSY searches. Imposing the isolation requirement induces logarithms $\alpha_s^n \ln^n\epsilon_\gamma$, with  $\epsilon_\gamma = E_{\rm iso}/E_\gamma$, into the perturbative computation and in the following we want to study their resummation.  

Already at the parton level, there are two mechanisms to produce a photon. In addition to the direct emission, one can produce an energetic quark which then fragments into a photon accompanied by a collinear quark. This second mechanism involves the fragmentation function, a non-perturbative object which needs to be extracted from data. In general, the two partonic contributions are not individually well-defined. At NLO, the direct production suffers from a divergence when the quark becomes collinear to the photon and this divergence is absorbed into the fragmentation function. The isolation cone suppresses fragmentation since it limits the amount of radiation which accompanies the photon. Indeed, Frixione has shown that one can modify the isolation criterion to eliminate fragmentation altogether \cite{Frixione:1998jh}.  For any angle $\delta<\delta_0$, where $\delta_0$ is the isolation cone angle, he imposes that the energy inside the cone of half-opening angle $\delta$ is smaller than 
\begin{equation}\label{eq:frixione}
 E_{\rm iso}(\delta) = \epsilon_\gamma E_\gamma \left(\frac{1 - \cos \delta}{1-\cos \delta_0}\right)^n\,,
\end{equation}
with $n>0$. Together with radiation collinear to the photon, this smooth-cone isolation eliminates the fragmentation contribution, which is centered at $\delta=0$. This simplifies the theoretical computations and is appealing because it eliminates the poorly known fragmentation function. Up to now, all NNLO computations of photon production \cite{Catani:2011qz,Campbell:2016yrh,Campbell:2016lzl} rely on the Frixione cone for isolation, while the result with a fixed cone is only known at NLO in the form of the {\sc JetPhox} code \cite{Catani:2002ny}. Due to the granularity of the calorimeter, a smooth criterion such as  \eqref{eq:frixione} cannot be directly implemented in experiments which therefore use fixed-cone isolation. To compare with experimental data, the NNLO results tune the parameters $\epsilon_\gamma$ and $n$ such that the NLO predictions using \eqref{eq:frixione} are numerically similar to fixed-cone computations including fragmentation. Below, we will derive such a parameter relation based on the analysis of soft radiation.

The  logarithms we want to study become large in the limit $\epsilon_\gamma \to 0$. In this limit the radiation inside the cone becomes very soft. It is well known that the emission of soft quarks is power suppressed and for this reason, fragmentation is a power suppressed effect for $\epsilon_\gamma \to 0$ which we do not need to consider. (The same holds true for threshold resummation studied in \cite{Becher:2009th}  and implemented into the numerical code {\sc PeTeR} \cite{Becher:2013vva}.)
As we discussed above, in the hadron collider case there are some interesting open issues and we therefore first derive a factorization theorem for  $e^+ e^-$. The kinematics is shown in Figure~\ref{fig:phiso_def}. One has hard partons outside the cone with energies of the order of the photon energy $E_\gamma$ and soft radiation inside the cone. This is precisely the situation captured by \eqref{sigbarefinal}, except that the soft region is now defined by the photon instead of the hard jets. Specializing the general formula to the photon case, we have
\begin{align}\label{masterFactorizationFormula}
\frac{\text{d}\sigma (\epsilon_\gamma,\delta_0)}{\text{d}x_\gamma}=\sum_{m=2}^{\infty}\left\langle \bm{\mathcal{H}}_{\gamma+m}\left(\{\underline{n}\},E_\gamma,Q,\delta_0\right)\otimes\bm{\mathcal{S}}_{m}\left(\{\underline{n}\}, \epsilon_\gamma \,E_\gamma, \delta_0\right)\right\rangle\text{,}
\end{align} 
where the photon energy is parameterized as $E_\gamma=x_\gamma \, Q/2$. The hard functions 
$\bm{\mathcal{H}}_{\gamma+m}$ are the squared amplitudes for the photon and $m$-parton process and are defined as in \eqref{eq:Hm}. In addition to the integrals over the energies of the $m$ partons at fixed  directions $\{\underline{n}\} = \{n_1,\cdots,n_m\}$ outside the isolation cone, they include an integral over the photon phase space together with its constraints (the energy $E_\gamma$ in the example \eqref{masterFactorizationFormula}). The soft functions are given by the Wilson line matrix element \eqref{eq:Sn} with the energy constraint applied to radiation inside the photon cone.

\begin{figure}[t!]
\begin{center}
\includegraphics[width=0.48\textwidth]{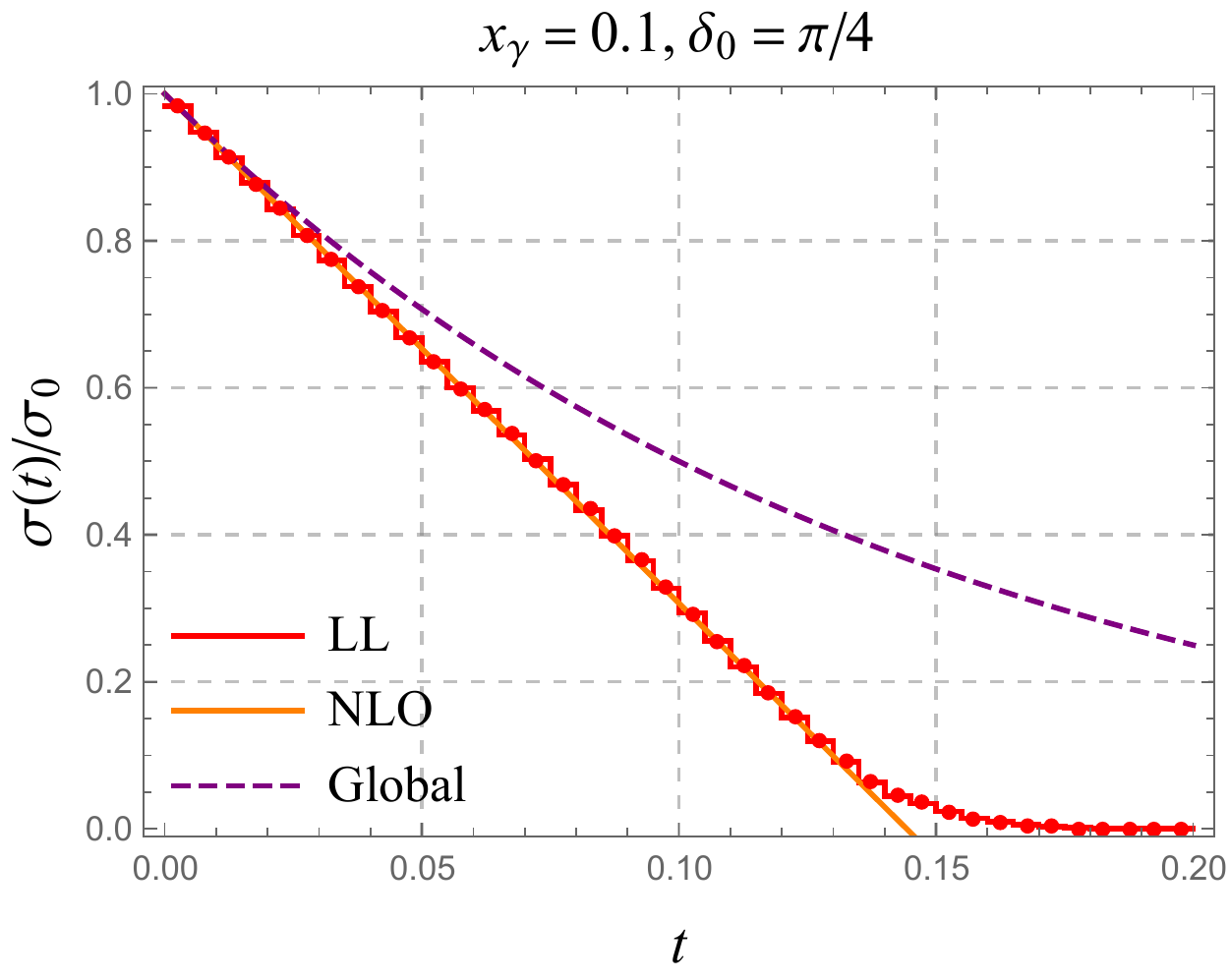}
\includegraphics[width=0.48\textwidth]{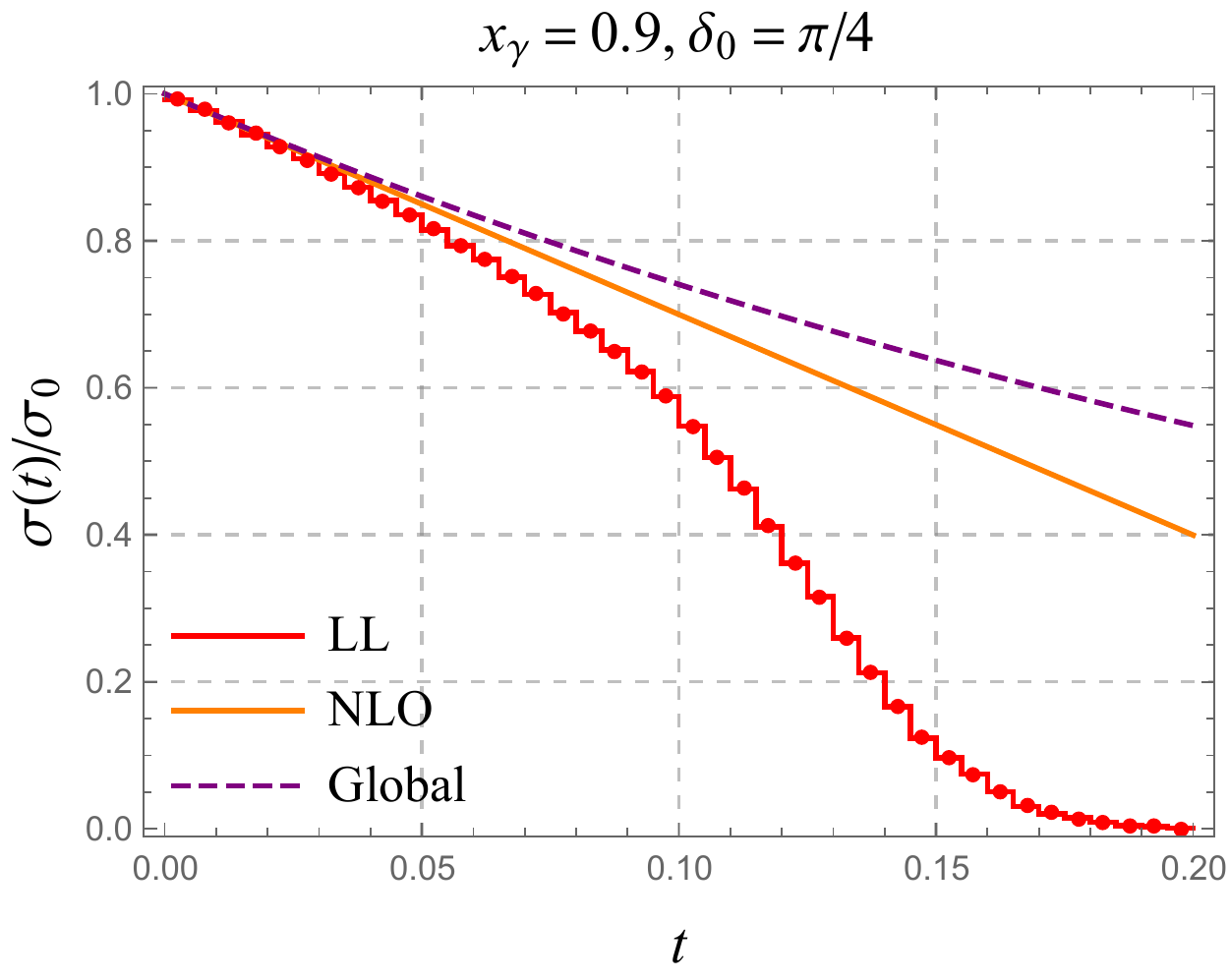}
\end{center}
\vspace{-0.5cm}
\caption{Effect of the isolation cut in $e^+e^- \to \gamma + X$. The plot shows a comparison of the resummed result (red line) with the one-loop contribution (orange line) and the global logarithms (dashed purple line).}
\label{fig:plxg}
\end{figure}

We will use the automated framework of the previous chapter to resum the large logarithms in the isolation cone cross section, but it is interesting to first analyze the NLO cross section analytically. The NLO correction to the soft function $\bm{\mathcal{S}}_2$ with two Wilson lines in $d=4-2\epsilon$ dimensions is given by the integral
\begin{align}\label{s2nlo}
\bm{\mathcal{S}}_2(\{n_1,n_2\},&\epsilon_\gamma \,E_\gamma,\delta,\epsilon) 
= \bm{1} - \bm{T}_1 \cdot \bm{T}_2 \,\,  g_{s}^2 \tilde{\mu}^{2\e} \int \frac{d^{d-1} k}{(2\pi)^{d-1}2\omega } \, \frac{2 \, n_1 \cdot n_2 \,}{n_1\cdot k \, n_2 \cdot k} \theta(E_{\rm iso} -  \omega)\,,
\end{align}
where $\omega = |\vec{k}|$ is the gluon energy. Note that the soft gluon can also be outside the isolation cone, but this part of the integration is scaleless and vanishes. Exactly the same integral is relevant for $\bm{\mathcal{S}}_m$, which involves a sum over all pairs of hard partons. In Appendix \ref{sec:NLOphoton}, the full computation of $\bm{\mathcal{S}}_2$ is performed analytically. To avoid technicalities and get a qualitative understanding, we will now perform an approximate computation. Since all hard partons are outside  while the soft gluon is inside the cone, the dipole factor is not singular. If the cone is narrow and the hard partons are not too close to the cone, we can approximate the gluon direction with the photon direction so that
\begin{equation}\label{eq:approx}
\frac{n_1 \cdot n_2 \,}{n_1\cdot k \, n_2 \cdot k} \approx
\frac{1}{\omega^2}\,\frac{n_1 \cdot n_2 \,}{n_1\cdot n_\gamma \, n_2 \cdot n_\gamma} =  \frac{1}{\omega^2} W_{12}^\gamma\,.
\end{equation}
The one-loop correction to the soft function then simplifies to
\begin{align}\label{softsimp}
\bm{\mathcal{S}}_2 \approx \bm{1} + C_F\,\bm{1}  \,  \frac{2g_{s}^2}{(2\pi)^{d-2}} \,  W_{12}^\gamma\, \int_0^\infty \frac{d\omega}{\omega} \, \left(\frac{\tilde{\mu}}{\omega}\right)^{2\epsilon} \int_{\rm cone} \frac{d\Omega}{4\pi} \, \theta(E_{\rm iso} -  \omega)\,.
\end{align}
For a fixed cone-energy $E_{\rm iso}$, the energy integration produces a divergence with an associated logarithm, which gets multiplied by the angular area of the cone, in line with the discussion in Section~\ref{sec:quali}. The situation is interesting for isolation cones because the logarithms are typically large (experiments often restrict the isolation energy to a few GeVs), while the area tends to be small. If we substitute $E_{\rm iso} \to E_{\rm iso}(\delta)$ from \eqref{eq:frixione} into \eqref{softsimp}, we can compute the soft function for the smooth-cone. In the approximation \eqref{eq:approx}, we find that the smooth-cone result is obtained from the fixed cone one-loop result using the substitution
\begin{equation}
\ln\frac{\epsilon_\gamma E_\gamma}{\mu} \;\;\longrightarrow\;\;  \ln\frac{\epsilon_\gamma e^{-n}E_\gamma }{\mu}\,.
\end{equation}
In other words, the smooth-cone isolation is more restrictive than fixed-cone isolation by a factor $e^{n}$. A computation such as \cite{Campbell:2016lzl} which uses smooth-cone isolation with $\epsilon_\gamma=0.1$ and $n=2$, therefore has the same size logarithms as a fixed-cone computation with $\epsilon_\gamma=0.01$. For photon energies of a few hundred GeVs, this indeed matches up with the fixed-cone isolation criterion 
\begin{equation}\label{eq:ATLAS}
E_T^{\rm iso} = 4.8 \, {\rm GeV} + 0.0042\, E^T_\gamma
\end{equation} 
used in the ATLAS analysis \cite{Aad:2016xcr}. ATLAS uses a cone of $R=0.4$ in the rapidity and azimuthal-angle plane. A particle is considered to be inside the cone (and therefore belongs to the ``out''-region), if $\Delta y^2+\Delta \phi^2< R^2$, where $\Delta y$ is the rapidity difference and $\Delta \phi$ the difference of the azimuthal angle between the particle and the photon.

As we discussed in Section  \ref{sec:quali} above, the two-loop non-global and global logarithms can cancel each other out and for photon isolation results displayed in Figure~\ref{fig:plxg}, this effect is quite pronounced. In this plot we consider $e^+e^- \to  \gamma + X$ with an isolation cone with half-angle $\delta_0=\pi/4$ and compare the resummed result with the one-loop logarithm and with the global contribution, which is given by the exponential of the one-loop logarithm. We observe that higher-order effects are quite small down to relatively low isolation energies which correspond to larger values of $t$ in the figure. Resumming the global logarithms leads to a much larger effect, which cancels after accounting also for the non-global contribution. By now there are many papers in the SCET literature which resum observables up to non-global contributions. This example demonstrates that such estimates of higher-order terms are not always reliable. In the present example this incomplete resummation leads to worse predictions than no resummation at all.

Finally, let us analyze photon isolation in hadronic collisions. Of course, in this case the same caveats apply that we discussed for gaps between jets: a full factorization analysis for hadronic collisions is not yet available. We will therefore again work in the large-$N_c$ limit and resum the leading logarithms captured by evolving the hard function from the scale $\mu_h \approx E_T^\gamma$ down to the soft scale $\mu_s  \approx E_T^{\rm iso}$. We need to evaluate the PDFs at the hard scale  $\mu_f =\mu_h$, as explained in the gaps-between-jets case. 
 
 \begin{figure}[t!]
\begin{center}
\begin{tabular}{ccc}
\includegraphics[width=0.47\textwidth]{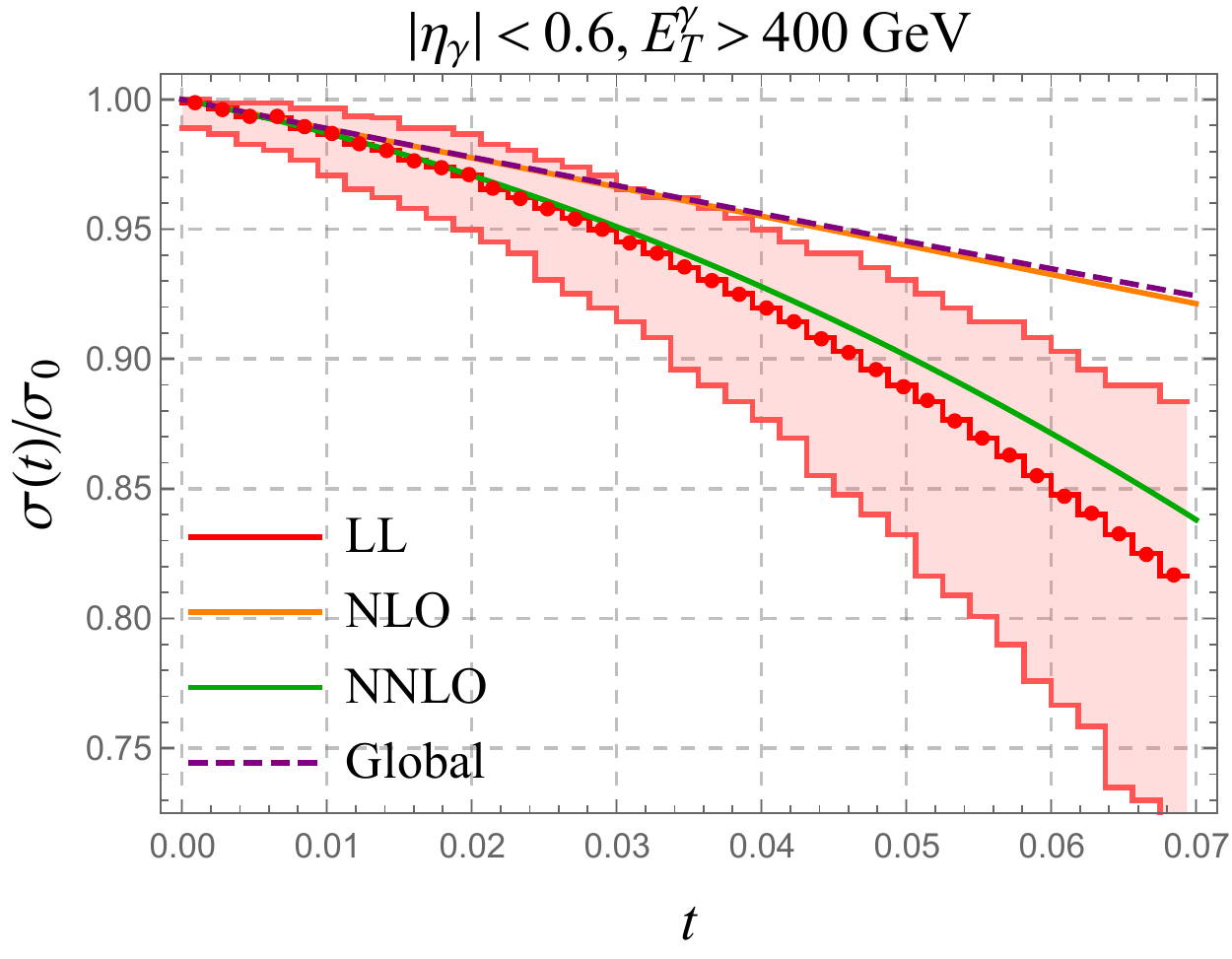} &&
\includegraphics[width=0.47\textwidth]{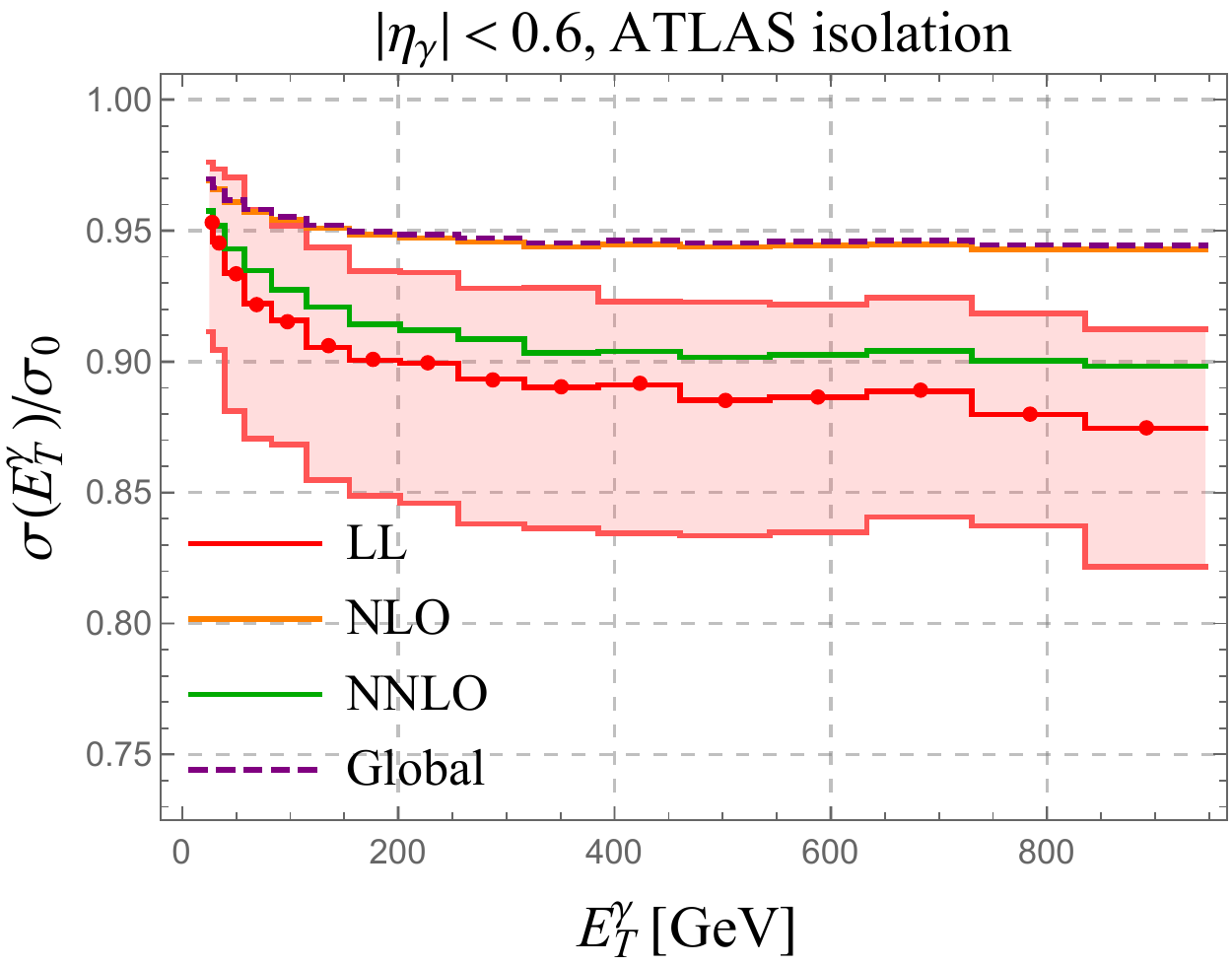} 
\end{tabular}
\end{center}
\vspace{-0.5cm}
\caption{Ratio of the $pp \to \gamma + X$ cross section with isolation to the inclusive one. Left: Ratio as a function of $t$ (or equivalently $\epsilon_\gamma$) for $E_T^\gamma > 400 \,{\rm GeV}$. Right: 
Ratio for the ATLAS isolation criterion \eqref{eq:ATLAS} as a function of $E_T^\gamma$. In both plots we show the resummed result as well as its NLO and NNLO expansions obtained using the approximation \eqref{eq:approx}. The red uncertainty bands are obtained by scale variations, see text.}
\label{fig:photon_pp}
\end{figure}
 
The small angular size $R$ of the veto region suppresses higher-order corrections and the overall effect of the isolation cone is therefore moderate. At the same time, the typical scale ratios $\epsilon_\gamma$ that arise in experimental measurements can be quite large. We have discussed in Section \ref{sec:quali} that the global logarithms  scale as $\alpha_s^n \,R^{2n}\, \ln^n(\epsilon_\gamma)$, while the non-global ones scale as $\alpha_s^n\, R^2  \,\ln^{n-1}(R)\, \ln^n(\epsilon_\gamma)$, since they involve only a single gluon in the veto region. For small $R$, the non-global logarithms completely dominate the cross section. In order to verify this, we extract large logarithms up to two-loop from our parton-shower code.  Explicitly, as is shown in \cite{Becher:2016mmh}, the first two coefficients in the expansion
\begin{equation}
\sigma(t)/\sigma_0 = 1 + \mathcal{S}^{(1)} t + \mathcal{S}^{(2)} t^2 + \dots
\end{equation}
in the shower time \eqref{eq:runt} take the form
\begin{align}
\mathcal{S}^{(1)} = &-4 N_c \int_\Omega {\bm 3}_{\rm out} W_{12}^3, \nno \\
\mathcal{S}^{(2)}= & \, \frac{(4N_c)^2}{2!}  \int_{\Omega}\Big[ -{\bm 3}_{\rm in}\,{\bm 4}_{\rm out}   \left(P_{12}^{34} - W_{12}^3\,W_{12}^4\right) + {\bm 3}_{\rm Out}\,{\bm 4}_{\rm Out}\,W_{12}^3\,W_{12}^4\Big],
\end{align}
where the subscript ${\rm ``in"}$ and ${\rm ``out"}$ refer to the radiation inside the jets (outside the isolation cone) and outside jets (inside the isolation cone), respectively. The coefficient of the one loop shower time $\mathcal{S}^{(1)}$ can be calculated using our MC simulation to generate a single emission along $n_3$ inside the cone (the ``out''-region). To calculate the non-global part of the two-loop coefficient  we approximate $n_4$ with the direction of the photon as we did in \eqref{eq:approx}, and end up with
\begin{align} 
\mathcal{S}_{\rm NG}^{(2)} \approx - \frac{(4N_c)^2}{2!} \Omega_{\rm cone} \int_\Omega {\bm 3}_{\rm in} W_{12}^3 \, \left(W_{13}^\gamma + W_{23}^\gamma -W_{12}^\gamma\right).
\end{align}
We then again use our MC simulation to generate vectors $n_3$ outside the cone (in the ``in''-region). Due to exponentiation, the global part of $\mathcal{S}^{(2)}$ is one-half of the one loop correction squared. Our results are shown in the left plot in Figure \ref{fig:photon_pp}, where we give evolution effects as a function of shower time as defined in \eqref{eq:runt}. The red line shows the LL resumed result, and the orange and green lines are one-loop- and two-loop-LL contributions, respectively. The dashed purple line corresponds to the naive exponentiation of one-loop results. To obtain the red error band, one first calculates $\tilde{\mu}=\mu_s(t)$ by inverting \eqref{eq:runt}. Varying this scale by a factor of two, one then obtains two values $t_{\rm low}=t(2\tilde{\mu})$ and $t_{\rm high}=t(\frac{1}{2}\tilde{\mu})$. The cross sections $\sigma(t_{\rm high})$  and $\sigma(t_{\rm low})$ are then used to define the uncertainty band. Clearly, there is a large difference between the one- and two-loop results, which is due to the $\ln R$ dependence of the NGLs which dominate the cross section. On the other hand, the difference between NNLO and the resummation is moderate.  In the right plot, we show resummation effects as function of photon transverse energy $E_T$ for the ATLAS \cite{Aad:2016xcr} isolation criterion \eqref{eq:ATLAS}. In this case, the red band is obtained by varying the soft scale by a factor two around the default value $\mu_s  = E_T^{\rm iso}$. Overall, resummation changes the NLO result for the isolation effects by about a factor of two. On the other hand, since higher-oder corrections beyond two loops are moderate, we don't anticipate large corrections to the NNLO computation in \cite{Campbell:2016lzl}. 

Until now we were focussing on logarithms of $\epsilon_\gamma$ arising in the limit of small isolation energy, while keeping the cone radius $R$ fixed. It is also interesting to keep $\epsilon_\gamma$ fixed and consider the limit of small $R$. That both limits are problematic for fixed-order computations was stressed already in \cite{Catani:2002ny} and the small $R$ case has been studied in detail in \cite{Catani:2013oma}, after it was realized that for narrow cones the NLO cross section with isolation \cite{Catani:2002ny} becomes larger than the inclusive one \cite{Gordon:1993qc}, which is of course unphysical. In \cite{Catani:2013oma}, the leading $\ln R$ terms were resummed using collinear factorization. It was found that the higher-order effects are moderate for $R\gtrsim 0.5$, but quickly become large for smaller cone radii. The paper \cite{Catani:2002ny} found that $\ln\epsilon_\gamma$ terms were moderate, but warned that the NLO computation could underestimate the overall effect. Our results in Figure \ref{fig:photon_pp} show that the nonglobal NNLO terms are as large as the NLO corrections, confirming this suspicion. 

Phenomenologically, the double limit $\epsilon_\gamma\to 0$ and $R \to 0$ is perhaps most relevant. We will now consider this situation, in which both types of logarithms are present. The relevant factorization analysis is quite similar to the one for the narrow-cone Sterman-Weinberg cross section \cite{Becher:2015hka,Becher:2016mmh}. In the following we will state and discuss the result; we refer the reader to \cite{Becher:2015hka,Becher:2016mmh} for more details regarding its derivation. Explicitly, for small $R\sim \delta_0$ the factorization formula \eqref{masterFactorizationFormula} turns into
\begin{align}\label{factorizationFormulaSmallCone}
\frac{\text{d}\sigma (\epsilon_\gamma,\delta_0)}{\text{d}E_\gamma}&= \frac{\text{d}\sigma^{\rm incl}_{\gamma+X}}{\text{d}E_\gamma}\nonumber\\
&\hspace{1cm}+\sum_{i=q,\bar{q},g} \int dz \frac{\text{d}\sigma_{i+X}}{\text{d}E_i}
\sum_{l=1}^{\infty}\left\langle  \bm{\mathcal{J}}_{\!\! i \to \gamma+l}\left(\{\underline{n}\}, \delta_0 \,E_\gamma,z \right)\otimes \bm{\mathcal{U}}_{l}\left(\{\underline{n}\}, \epsilon_\gamma\, \delta_0 \,E_\gamma\right)\right\rangle\text{.}
\end{align} 
In this formula, the first term on the right-hand side is the direct photon production cross section without photon isolation and without fragmentation. This term is obtained when considering soft radiation at paramatrically large angles $\delta\gg \delta_0$ for which one can ignore the narrow cone. Doing so renders the soft functions trivial and one can integrate over the directions of the hard partons, which yields the cross section $\sigma_{\gamma+X}$. The (perturbative) fragmentation contribution is part of the second term which describes the inclusive production of a parton $i$ along the photon direction, which then fragments into a photon plus soft hadronic radiation along the direction of the small isolation cone and energetic radiation immediately outside the cone. More precisely, the term $\sigma_{i+X}$ in the second line denotes the inclusive cross section for producing a parton $i$ with energy $E_i$ and momentum $p_i$ along the direction $n^\mu=n_\gamma^\mu$ of the photon, and the jet functions 
\begin{align}\label{jetfun}
   \frac{n\!\!\!/}{2}\,\bm{\mathcal{J}}_{\!\! i \to \gamma+ l}(\{\underline{n}\},\delta_0 \, E_\gamma, & z) 
   = \sum_{\rm spins}  \prod_{j=1}^l \int \! \frac{dE_j \,E_j^{d-3} }{(2\pi)^{d-2}} \,
    |\mathcal{M}_l(p_i;\{p_\gamma,\underline{p}\})\rangle \langle\mathcal{M}_l(p_i;\{p_\gamma,\underline{p}\})| \nonumber \\
  & \times  2\,(2\pi)^{d-1}\, \delta(2\,(1-z)\,E_i-\bar{n}\cdot p_{X_c})\,\delta^{(d-2)}(p_{X_c}^\perp)\, 
    {\Theta }^{n}_{\rm cone}\!\left(\left\{\underline{p}\right\}\right) .
\end{align}
describe the fragmentation of this parton into a photon with energy $E_\gamma= z E_i$  and $l$ additional energetic partons outside the cone, as enforced by 
the theta function ${\Theta }^{n}_{\rm cone}$ in their definition. The function $\bm{\mathcal{U}}_{l}$ describes soft radiation collinear to the isolation cone and consists of $l$ Wilson lines along the energetic partons plus one additional Wilson line along the light-cone direction $\bar{n}^\mu$ conjugate to the one of the photon direction. More details on this collinear and soft (or ``coft'') mode can be found in \cite{Becher:2015hka,Becher:2016mmh}. Its most important property is that the typical invariant mass of this type of radiation has the low value $\Lambda_{\rm coft} = \delta_0 \,\epsilon_\gamma\, E_\gamma$, precisely because it is both soft and collinear. In appendix \ref{sec:NarrowConePhoton}, we will evaluate the narrow-cone isolation cross section at leading order and verify that the QCD result maps onto the factorization theorem \eqref{factorizationFormulaSmallCone}. 

We note that the two terms in  \eqref{factorizationFormulaSmallCone} are not separately finite: the partonic cross sections and jet functions must be viewed as Wilson coefficients of the effective theory, which must be renormalized. To perform the resummation of the large logarithms, one has to solve the associated RG-evolution equations and first evolve from the hard scale $\mu_h \sim E_\gamma$ down to the jet scale $\mu_j \sim \delta_0 E_\gamma$ and finally to the coft scale $\mu\sim  \delta_0 \,\epsilon_\gamma\, E_\gamma$. 
As we discussed above, the quantity $\sum_l \langle \bm{\mathcal{J}}_{\!\! i \to \gamma+l} \otimes \bm{\mathcal{U}}_{l} \rangle$ 
describes the fragmentation of the parton $i$ into a photon plus soft and collinear radiation. It has exactly the same scale dependence as the standard photon fragmentation function, see \cite{Gluck:1992zx,Bourhis:1997yu}. The first step of RG evolution, which generates the logarithms of $R$ through the ratio $\mu_j/\mu_h$, is thus governed by the standard RG evolution of the fragmentation function. Logarithms of $\epsilon_\gamma$ are only generated in the second step, via the evolution from $\mu_j \sim \delta_0 E_\gamma$ down to $\mu\sim  \delta_0 \,\epsilon_\gamma\, E_\gamma$. We postpone a study of the numerical size of the $\ln R$ terms to future work.

\subsection{Jet-veto cross sections}

Rejecting events with hard jets can be important to make precise measurements at hadron colliders. An example is the process $p\,p \to W^+ W^-$ at the LHC, where the veto is used to reduce the background from top-quark pair production with subsequent $t\to b \, l \,  \nu$ decay. The cut used by ATLAS  rejects events with jets of $p_T^J > p_T^{\rm veto} = 25$~GeV for $|\eta^J|<4.5$ \cite{ATLAS:2012mec}, while CMS imposes $ p_T^{\rm veto} = 30$~GeV for $|\eta^J|<5$  \cite{Chatrchyan:2013yaa}. The jet-veto cut introduces logarithms $\ln(p_T^{\rm veto}/m_H)$, which can spoil the convergence of perturbative calculations. Much work has been carried out to resum these large logarithms \cite{Banfi:2012yh,Becher:2012qa,Banfi:2012jm,Becher:2013xia,Stewart:2013faa}. The resummation at NNLL+NLO accuracy has been automated for the production of an arbitrary final state with massive colorless particles within the MadGraph5$\_$aMC@NLO framework \cite{Becher:2014aya}. 

The jet-veto cross section is a non-global observable, since the cross section becomes fully inclusive in the large rapidity region near the beams, because the veto can only be imposed where detectors are present.  Of course, this problem affects all hadron collider observables and in particular also hadronic event shapes. The NGLs in the jet-veto cross section have never been resummed, but \cite{Banfi:2012yh} has analyzed the rapidity cut dependence in fixed order and by using parton showers, and concluded that it was small.  The paper \cite{Hornig:2017pud} pointed out that the non-global effects are power suppressed for the kinematic cuts used at the LHC. In order to explain this power suppression effects, let us first define two expansion parameters
\begin{align}
\beta = p_T^{\rm veto}/Q,~~~ \delta = e^{-\eta_c},
\end{align}
where $\eta_c$ is the rapidity cut, and $Q$ represents the hard scale for this process. E.g. for $W^+ W^-$ production it is the invariant mass of the  electroweak final state  $Q= M_{W^+W^-}$. 

For jet-vetoed cross section at the LHC, the hierarchy between the two parameters is $\beta\sim 0.1 \gg \delta \sim 0.01$. Analyzing which momentum regions are relevant, one finds that collinear modes contributing to jet-veto resummation have light-cone components $(n\cdot p_c, \bar{n}\cdot p_c, p^\perp)$ scaling as $Q(\beta^2, 1, \beta)$, where $n^\mu$ and $\bar{n}^\mu$ are light-cone vectors along the beams. The typical rapidity of these particles is much smaller than the cut $\eta_c \sim 5$ used at the LHC.  Contributions sensitive to the rapidity cut $\eta_{\rm cut}$ are therefore power suppressed by $\delta/\beta$. This parametric suppression is consistent with the small size of the fixed-order corrections computed in \cite{Banfi:2012yh}. 

One can also consider the opposite hierarchy  $\beta \ll \delta \ll 1$, as analyzed in \cite{Hornig:2017pud}. At LHC energies, the low $p_T^{\rm veto}$ scale related with $\beta$ would  be non-perturbative in this situation, so it is currently only of theoretical interest. To capture the physics in the low-energy region one needs modes with the same scaling behavior as the coft mode introduced in \cite{Becher:2015hka}. The paper \cite{Hornig:2017pud} analyzed the factorization for rapidity-dependent jet-veto cross sections but their analysis was restricted to global logarithms. We recently developed the necessary framework to deal with soft-recoil sensitive non-global observables in \cite{Becher:2017nof} and it would be interesting to derive the full formula in our framework. 

\section{Conclusion}\label{conclusion}

In this paper, we have used RG methods in effective field theory to obtain a parton shower for the resummation of large logarithms in non-global observables. Our result provides an explicit example of a parton-shower equation derived from first principles which can be systematically improved. At LL level in the large-$N_c$ limit, our shower is equivalent to the Dasgupta-Salam dipole shower. We have implemented it and have interfaced it with {\sc MadGraph5\Q{_}aMC@NLO} to obtain a flexible framework to perform resummations. The tree-level generator is used to produce a LHE file containing the kinematic configuration and color structure of the hard partons. This information is then passed to the shower to perform the RG evolution to lower scales. 

With this method we have investigated gap fractions in dijet production and isolation cone cross sections. We find that non-global contributions are especially important when the veto region is small, because the higher-order global contributions are suppressed by higher powers of the size of the veto region, while this suppression is absent for the non-global terms. We observe that the LL predictions suffer from large uncertainties, and it will be important to extend the resummation to higher accuracy in the future. In addition, there are also several other issues, which can and should be studied already at the leading logarithmic level, such as the role of momentum conservation to reduce power corrections and the resummation of collinear logarithms. For exclusive jet cross sections, we have shown in earlier work how the collinear logarithms arising for small jet radius can be resummed, and in the present work we have extended the relevant factorization to small isolation cones. As in the case of small-radius jets, we find that momentum modes are relevant, which are both soft and collinear to the cone.

To resum next-to-leading logarithms, one needs higher-order corrections to the anomalous dimension matrix and the matching coefficients. Specifically, one will need to include i.) the one-loop soft functions $\bm{\mathcal{S}}_{m}$ for any $m$, ii.) the one-loop correction to the Born-level hard function $\bm{\mathcal{H}}_k$ and the tree-level result for $\bm{\mathcal{H}}_{k+1}$, the hard function with one additional emission. In addition, one also needs iii.) the two-loop anomalous dimension. In earlier papers, we have computed i.) and ii.) for specific processes and iii.) should have a close relation to the result of Caron-Huot in the density matrix  formalism \cite{Caron-Huot:2015bja}. While our RG framework makes it clear which ingredients are necessary to improve the logarithmic accuracy, it will likely be nontrivial to implement these into a MC framework similar to the one we employed at LL. Nevertheless, it is important to pursue this line of research, not only to reduce the uncertainties in the observables studied here, but also because it can provide a first example of a parton shower with higher logarithmic accuracy.

Our shower code is currently restricted to the large-$N_c$ limit, but it would be interesting to go beyond this approximation, especially for hadron-collider processes, where contributions from Glauber phases arise at finite $N_c$. Without accounting for these in the low-energy theory, the factorization theorem would not be RG invariant because the double-logarithmic evolution of the hard functions, which produces the ``super-leading'' logarithms, could not be matched by the evolution of the operators in the low-energy theory. A detailed discussion of these effects will be given in a forthcoming paper.

\begin{acknowledgments}	
The research of T.B.\ is supported by the Swiss National Science Foundation (SNF) under grant CRSII2\_160814. The authors would like to thank Jeppe Andersen, Helen Brooks, Keith Hamilton, Johannes Michel, Pier Francesco Monni,  Matthias Neubert, Simon Pl\"atzer, Emanuele Re and Gavin Salam for useful discussions and the Munich Institute for Astro- and Particle Physics (MIAPP) of the DFG cluster of excellence "Origin and Structure of the Universe" for hospitality and support.
\end{acknowledgments}  

\appendix

\section{\boldmath Angular integration with a collinear cutoff \unboldmath\label{sec:labcone}}

With a collinear cutoff $\lambda$ the angular integration in the anomalous dimensions $\bm{V}_m$ and  $\bm{R}_m$ in \eqref{eq:oneLoopRG} takes form
\begin{equation}\label{eq:labcut}
I(\lambda, n_i, n_j)=\int \frac{d\Omega(n_l)}{4\pi} \frac{n_i \cdot n_j}{ n_i \cdot n_l \, n_l \cdot n_j} \theta( n_l \cdot n_i - \lambda^2 ) \theta(n_l \cdot n_j - \lambda^2)\,.
\end{equation}
The cutoff amounts to putting small cones around the emitting partons to avoid the collinear singularity. In the lab frame any vector $n_l$ can be parametrised as 
\begin{align}
n_l &= (1, {\rm sech\,}y_l\sin\phi_l, {\rm sech\,}y_l\cos\phi_l,\tanh y_l)\,.
\end{align}
In order to compute \eqref{eq:labcut}, we transform the integration into the Center-Of-Mass (COM) frame of $n_i$ and $n_j$, where it takes the form
\begin{align}\label{comint}
I(\lambda, M)= \int_{-\infty}^{\infty} d  \hat y_l \int_0^{2\pi} \frac{ d  \hat \phi_l}{2\pi} \, \theta\!\left[  \frac{M^2(1-\tanh \hat  y_l)}{4(1-\beta \cos\hat\phi_l \, {\rm sech\,} \hat y_l)}  - \lambda^2 \right] \theta\!\left[  \frac{M^2(1+\tanh  \hat y_l)}{4(1-\beta \cos \hat  \phi_l \, {\rm sech\,} \hat  y_l)}  - \lambda^2 \right]\, .
\end{align}
Here $M^2=2\,n_i\cdot n_j$ is the invariant mass of the $n_i$ and $n_j$ dipole, and $\beta=\sqrt{1-M^2/4}$.  The new integration variables $\hat y_l$ and $\hat \phi_l$ are the rapidity and azimuthal angle of the emission in the COM frame.  The components $n_l^\mu =(1,n_x,n_y,n_z)$  in the lab frame can be expressed in terms of $\hat y_l$ and $\hat \phi_l$ as
\begin{align}
n_x &=  \frac{E_l}{M} \Bigg[ \left( 1 - \frac{\cos\hat\phi_l \, {\rm sech\,}\hat y_l}{\beta} \right) \big( \, {\rm sech\,} y_i  \sin\phi_i + \, {\rm sech\,} y_j \sin\phi_j \big) + \tanh \hat y_l \big( \, {\rm sech\,} y_i  \sin\phi_i \nno \\
&~~~~- \, {\rm sech\,} y_j \sin\phi_j \big)  + \frac{\, {\rm sech\,} \hat y_l  \sin\hat \phi_l}{\beta} \left( \cos\phi_i \, {\rm sech\,} y_i  \tanh y_j -  \cos\phi_j \, {\rm sech\,} y_j  \tanh y_i  \right)     \Bigg]\,, \nno \\
n_y &=n_x(\sin\phi_{i,j}\to \cos\phi_{i,j}\,, \cos\phi_{i,j}\to -\sin\phi_{i,j})\,,\nno \\
n_z &=  \frac{E_l}{M} \Bigg[ \left( 1 - \frac{\cos\hat\phi_l \, {\rm sech\,}\hat y_l}{\beta} \right) \left( \tanh y_i + \tanh y_j \right) + \tanh\hat y_l \left( \tanh y_i - \tanh y_j \right)  \nno \\
&~~~~+ \frac{1}{\beta} \, {\rm sech\,} y_i \, {\rm sech\,} y_j \, {\rm sech\,}\hat y_l  \sin(\phi_i - \phi_j) \sin\hat\phi_l   \Bigg]\,. 
\end{align}
with $M/E_l = 2-2\beta \cos\hat\phi_l \, {\rm sech\,} \hat y_l$. The result for the components will be useful for the phase-space generation for the real emissions. To obtain the virtual corrections, we now evaluate (\ref{comint}). As long as the two cones around $n_i$ and $n_j$ do not touch each other, i.e.\ for $M^2 > 8 \lambda^2-4\lambda^4$, the integration constraints implemented by the $\theta$-function in (\ref{comint}) reduce to
\begin{align}\label{eq:Icms}
I_1(\lambda, M)= \int_0^{2\pi} \frac{ d\hat \phi_l}{2\pi} \int_{-y_{\rm max}(\hat{\phi}_l)}^{y_{\rm max}(\hat{\phi}_l)} d \hat y_l\,,
\end{align}
with
\begin{align}
y_{\rm max}(\hat{\phi}_l) &= \ln\!\left( \beta\cos\hat{\phi}_l  + \sqrt{\alpha+ \beta^2\cos^2(\hat{\phi}_l) } \right)\,,
\end{align}
where $\alpha = (M^2-2\lambda^2)/(2\lambda^2)$. Performing these integrations, one obtains the analytical result
\begin{align}
 I_1(\lambda, M)= \ln\!\left(\frac{M^2}{2\lambda^2} - 1\right).\end{align}
In the region $2\lambda^2 < M^2  < 8 \lambda^2-4\lambda^4$ the integration boundary can be simplified to
\begin{align}
I_2(\lambda, M)= \int_0^{\delta} \frac{ d \hat \phi_l}{\pi} \int_{-y_{\rm max}(\hat{\phi}_l)}^{y_{\rm max}(\hat{\phi}_l)} d \hat y_l\,,
\end{align}
with $\cos\delta=(1-\alpha)/(2\beta)$. Because the two cones overlap, the azimuthal angle integration is now restricted. After performing integration by parts, $I_2$ can be reduced to a one-dimensional elliptic integral
\begin{align}
I_2(\lambda, M)= \frac{2\beta}{\pi}\int_0^{\delta} d\hat\phi_l \frac{\hat\phi_l \sin\hat\phi_l}{\sqrt{\alpha + \beta^2 \cos^2\hat\phi_l}}\,.
\end{align}
Since we do not have an analytical result, we use numerical interpolation for $I_2(\lambda, M)$ in our parton-shower code. 

The form of the collinear cutoff is of course not unique. A simpler form of the virtual integral is obtained by imposing the cutoff in the COM frame by putting a cut on $\hat y$. The angular integration then reads
\begin{align}\label{eq:DScut}
\tilde{I}(\lambda, M)= \int_0^{2\pi} \frac{ d \hat \phi_l}{2\pi} \int_{-\tilde y_{\rm max}}^{\tilde y_{\rm max}} d \hat y_l\,,
\end{align}
with $\tilde y_{\rm max} =y_{\rm max}(0)= \ln\!\left( \beta + \sqrt{\alpha+ \beta^2}\right)$, so that $\tilde I(\lambda, M)= 2 \tilde y_{\rm max}$. This regularization scheme was used by Dasgupta and Salam \cite{Dasgupta:2001sh}. We will compare MC results based on the two cutoff schemes \eqref{eq:labcut} and \eqref{eq:DScut} in Appendix \ref{sec:MCdetail}.

\section{\boldmath Details of the MC algorithm \unboldmath}\label{sec:MCdetail}
In this appendix we will describe the MC algorithm in detail, working with the interjet energy flow in $e^+e^-$ for concreteness. For this observable, the lowest multiplicity hard function has two energetic partons along back-to-back directions $n_1$ and $n_2$. We can thus set $k=2$ in the equations in Section \ref{Rgrun}.
For more complicated observables, such as hadron collider dijet events, we start with $k>2$ partons, whose directions are read from an event file produced by the {\sc MadGraph} tree-level generator. The tree-level generator also assigns large-$N_c$ dipole color structure to each event, which we use as the starting point of our shower.
 
We will first spell out the algorithm and then show how it arises from the iterative solution of the RG-evolution equation of the hard functions in \eqref{eq:iterRG}. The basic ingredient of the MC algorithm is a list of events. Each event $E$ occurs at a time $t$, has a weight $w$ and contains a list of $m$ vectors $\{n_1,n_{i_1},\dots,n_{i_{m-2}}, n_2\}$. This list defines the color dipoles of the events, which are given by neighbouring pairs of vectors so that the associated virtual correction is
\begin{equation}
V_E = V_{1i_{1}} + V_{i_{1} i_{2}}+\dots + V_{i_{m-2} 2}\,,
\end{equation}
with
\begin{equation}
V_{ij} = \int\! \frac{d\Omega(n_l)}{4\pi}\, R_{ij}^l\,.
\end{equation}
The integrand is the real-emission matrix element
\begin{equation}
 R_{ij}^l = 4 \, N_c W_{ij}^l  \, \theta( n_l \cdot n_i - \lambda^2 ) \theta(n_l \cdot n_j - \lambda^2)\, .
\end{equation}
The angular integration in the presence of a collinear cutoff $\lambda$ was discussed in detail in  Appendix \ref{sec:labcone}. Note that the quantity $V_{ij}$ defined here is positive, while $\bm{V}_m$ in \eqref{eq:virtual} is negative.

The MC algorithm described in the following produces a histogram of $V_{12}\,\sigma_{\rm veto}(\Omega_0,t)/\sigma_0$.  To get the gap fraction that one has to divide the result by $V_{12}$, the virtual correction associated with the original dipole. The algorithm involves the following steps:
\begin{enumerate}
\item  \label{one}Start at shower time $t=0$ from an initial event with vectors $\{  n_1,  n_2 \}$ and weight $w=1$.
\item  \label{two} Generate a random time step $\Delta t$ according to the probability distribution $\mathcal{P}_E(t)= V_E \exp(-V_E \Delta t)$, and insert the event weight $w$ into the histogram at time $t+\Delta t$.
\item \label{three} Choose a dipole associated with a pair of neighbouring vectors $n_i$ and $n_j$ in $E$ with probability $V_{ij}/V_E$. Generate a new random vector $n_k$ and multiply the weight by the factor $R_{ij}^k/V_{ij}$, expressed in the random variables chosen to generate the direction of the new vector $n_k$, see \eqref{weight} below. 
\item \label{four} If $n_k$ is inside the veto region, go to  Step \ref{one} and start a new event, otherwise add this new vector into $E'=\{n_1,\cdots,n_{i},n_k,n_j,\cdots,n_2\}$, multiply the weight by a factor $V_E/V_{E'}$ and return to Step \eqref{two}. 
\end{enumerate}
To keep the weights $w$ close to one, one works in the COM variables $\hat{y}_k$ and $\hat{\phi}_k$ introduced in Appendix \ref{sec:labcone} to generate the direction of the new parton. In the dipole COM frame the integrand becomes trivial in these variables, see \eqref{eq:Icms}. However, with a lab-frame cut, the integration boundary   $y_{\rm max}(\hat{\phi}_k)$ in the rapidity integration depends on $\hat{\phi}_k$. Mapping the boundary to a square introduces a weight factor
\begin{equation}\label{weight}
w =\frac{2 y_{\rm max}(\hat{\phi}_k) \,\phi_{\rm max}}{V_{ij}/(4N_c)}\,.
\end{equation}
If one follows Dasgupta and Salam \cite{Dasgupta:2001sh} and introduces the collinear cutoff in the COM frame, the integration region is rectangular and $w=1$. A second advantage of this cutoff is that the weight factor in Step \ref{four} is always smaller than one, $V_E/V_{E'}< 1$. One can thus implement this factor by throwing away the event in Step \ref{four} with probability $V_E/V_{E'}$. Once this is done, one has unweighted events. In contrast, with a lab-cone cutoff a small fraction of events has  $V_E/V_{E'}> 1$. 

\begin{figure}[t!]
\begin{center}
\hspace{-1cm}
\begin{overpic}[scale=0.8]{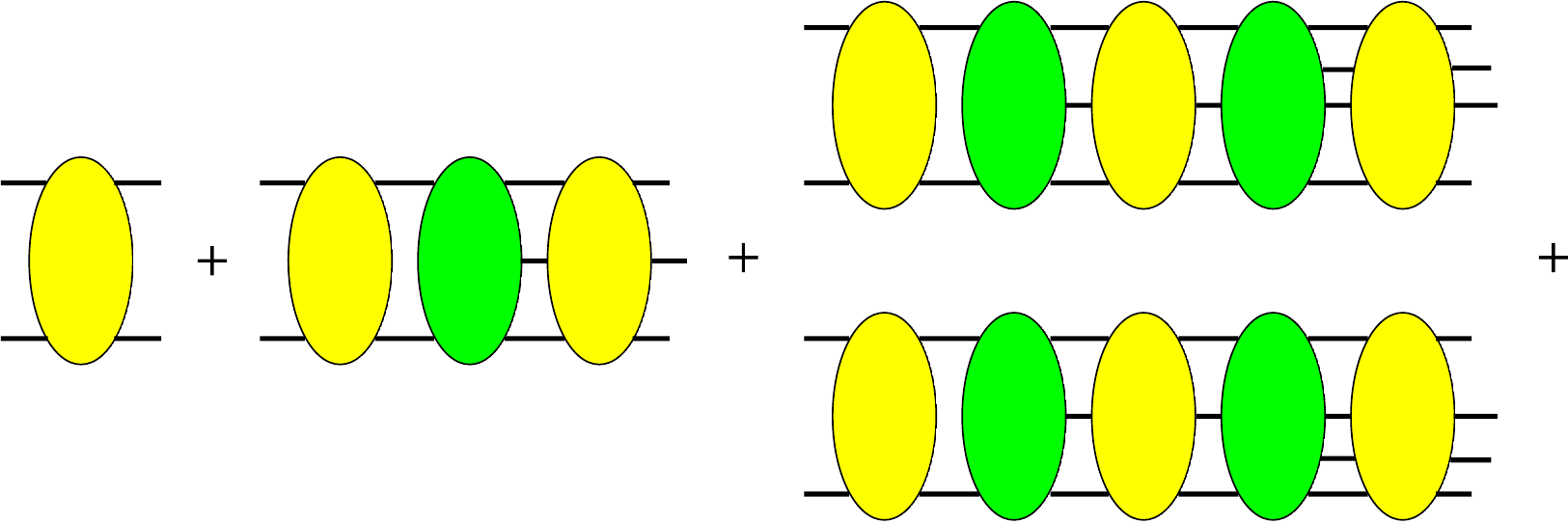}
\put(4,15.5){$V_2$}
\put(20,15.5){$V_2$}
\put(28.5,15.5){$R_2$}
\put(37,15.5){$V_3$}
\put(55,26){$V_2$}
\put(63,26){$R_2$}
\put(71.5,26){$V_3$}
\put(78.5,26){$R_3^{(1)}$}
\put(87,26){$V_4^{(1)}$}

\put(55,6){$V_2$}
\put(63,6){$R_2$}
\put(71.5,6){$V_3$}
\put(78.5,6){$R_3^{(2)}$}
\put(87,6){$V_4^{(2)}$}

\put(100,16){$~~\cdots$}

\put(-3,23){$n_1$}
\put(-3,9){$n_2$}
\put(10,23){$n_1$}
\put(10,9){$n_2$}
\put(4,0){$\mathcal{H}_2$}

\put(15,23){$n_1$}
\put(15,9){$n_2$}
\put(42,23){$n_1$}
\put(42,18){$n_3$}
\put(42,9){$n_2$}
\put(28.5,0){$\mathcal{H}_3$}

\put(49,32.5){$n_1$}
\put(49,23){$n_2$}
\put(49,13){$n_1$}
\put(49,3){$n_2$}
\put(95,32.5){$n_1$}
\put(95.5,29){$n_4$}
\put(96,26){$n_3$}
\put(95,21){$n_2$}
\put(94.5,12){$n_1$}
\put(96,7){$n_3$}
\put(96,3){$n_4$}
\put(95,0){$n_2$}

\put(72,38){$\mathcal{H}_4^{(1)}$}
\put(72,-7){$\mathcal{H}_4^{(2)}$}

\end{overpic}
\end{center}
\vspace{.5cm}
\caption{Diagrammatic representation of the lowest hard functions contributing to (\ref{xsec_exp}). \label{fig:evolve_fig}}
\end{figure}
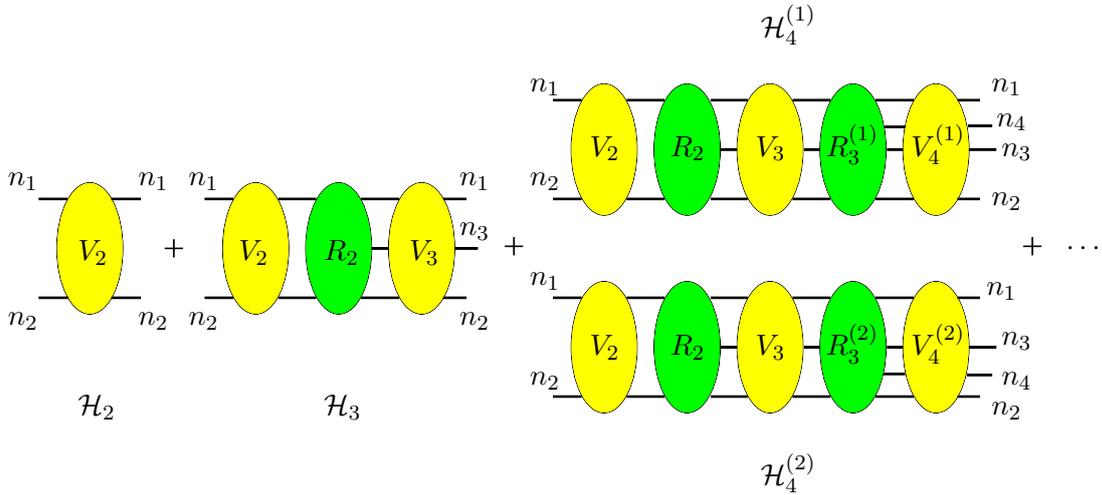

To derive the above MC algorithm, we  rewrite RG evolution solution \eqref{eq:iterRG} in a form which makes the four steps of the algorithm manifest. According to \eqref{eq:sigmaLL}, after evolving the hard functions to the soft scale $Q_0$, the veto cross section takes the form
\begin{align}\label{xsec_exp}
\widehat{\sigma}_{\rm veto}(\Omega_0,t)  =  \frac{V_{12}}{\sigma_0} \sigma_{\rm veto}(\Omega_0,t) = \mathcal{\widehat{H}}_2(t) + \int \frac{d\Omega_3}{4\pi} \mathcal{\widehat{H}}_3(t,n_3) + \int \frac{d\Omega_3}{4\pi}\frac{d\Omega_4}{4\pi} \mathcal{\widehat{H}}_4(t,n_3,n_4) + \cdots,
\end{align}
where the hat indicates the factor $V_{12}/\sigma_0$ by which we have multiplied the cross section and the hard functions $\mathcal{H}_m$ in order to work with the same normalization as the MC simulation. In Figure~\ref{fig:evolve_fig} we show their diagrammatic representations. The first term $\mathcal{\widehat{H}}_2$ represents no emission down to the veto scale $Q_0$, corresponding to shower-time evolution from $0$ to $t$. This purely virtual contribution takes the form
\begin{align}
\mathcal{\widehat{H}}_2(t)= \mathcal{P}_2(t) = V_{12} \, e^{-t \,V_{12}} \,.
\end{align}

As shown in Figure~\ref{fig:evolve_fig}, the second term $\mathcal{\widehat{H}}_3$ corresponds to a situation, where no emission occurs until the shower evolves to $t'$, at which time a new parton is emitted along the direction $n_3$, after which the system evolves without further emissions to $t$. This yields the expression
\begin{align}\label{eq:H3}
\mathcal{\widehat{H}}_3(t) & =  \int_0^t \!dt'\, \mathcal{\widehat{H}}_2(t') \, R_{12}^3 \, e^{-(t-t')V_3},
\end{align}
where the new virtual part  is $V_3=V_{13} + V_{32}$. We  now rewrite \eqref{eq:H3} in terms of factors which can be viewed as probabilities
\begin{align}\label{eq:H3r}
\mathcal{\widehat{H}}_3(t) & = \int_0^t \!dt'\, \mathcal{P}_2(\Delta t) \, \frac{R_{12}^3}{V_2} \, \frac{V_2}{V_3} \,  \mathcal{P}_3(\Delta t') \, , 
\end{align}
with $\Delta t = t'$ and $\Delta t' = t-t'$. To get an emission probability, we normalized the angular integral to $V_2$. Introducing the probability $\mathcal{P}_3$ for the second time step, we are then left with a factor $\frac{V_2}{V_3}$ which arises as a weight in Step \ref{four} of the algorithm.

Starting from $\mathcal{\widehat{H}}_4$, each hard function is a sum of several terms, which correspond to the different dipoles which can emit. Specifically, for $\mathcal{\widehat{H}}_4$ we have
\begin{align}\label{eq:H4}
\mathcal{\widehat{H}}_4(t) = \mathcal{\widehat{H}}_4^{(1)}(t) + \mathcal{\widehat{H}}_4^{(2)}(t)\,, 
\end{align}
where $\mathcal{\widehat{H}}_4^{(1)} $ corresponds to inserting a new parton into the dipole formed by $n_1$ and $n_3$ and has  the form
\begin{align}
\mathcal{\widehat{H}}_4^{(1)}(t) = R_{13}^4 \int_0^t \!dt'' \,\mathcal{\widehat{H}}_3(t'')  \, e^{-(t-t'')V_4^{(1)}}, 
\end{align}
with $V_{4}^{(1)} = V_{14} + V_{43} + V_{32}$. The second term $\mathcal{H}_4^{(2)} $ arises from inserting a new parton between $n_3$ and $n_2$. We rewrite \eqref{eq:H4} in the same form as \eqref{eq:H3r} and get
\begin{align}
\mathcal{\widehat{H}}_4(t) =  \int_0^t \!dt''\,\mathcal{\widehat{H}}_3(t'')  \, \left[  \frac{R_{13}^4}{V_{13}} \frac{V_{13}}{V_3} \frac{V_3}{V_{4}^{(1)}} \mathcal{P}_4^{(1)}(\Delta t'')  + \frac{R_{32}^4}{V_{32}} \frac{V_{32}}{V_3} \frac{V_3}{V_{4}^{(2)}} \mathcal{P}_4^{(2)}(\Delta t'') \right]\,,
\end{align} 
where $\Delta t''=t-t''$. Compared to \eqref{eq:H3r} we encounter additional factors $V_{13}/V_3$ and $V_{32}/V_3$, which represent the probability of choosing one of the two dipoles. These factors are implemented in Step \eqref{three} of the MC algorithm. No additional complications arise at higher multiplicities. 

\begin{figure}[t]
 \centering
 \hspace{-.5cm}
 \includegraphics[width=0.47\textwidth]{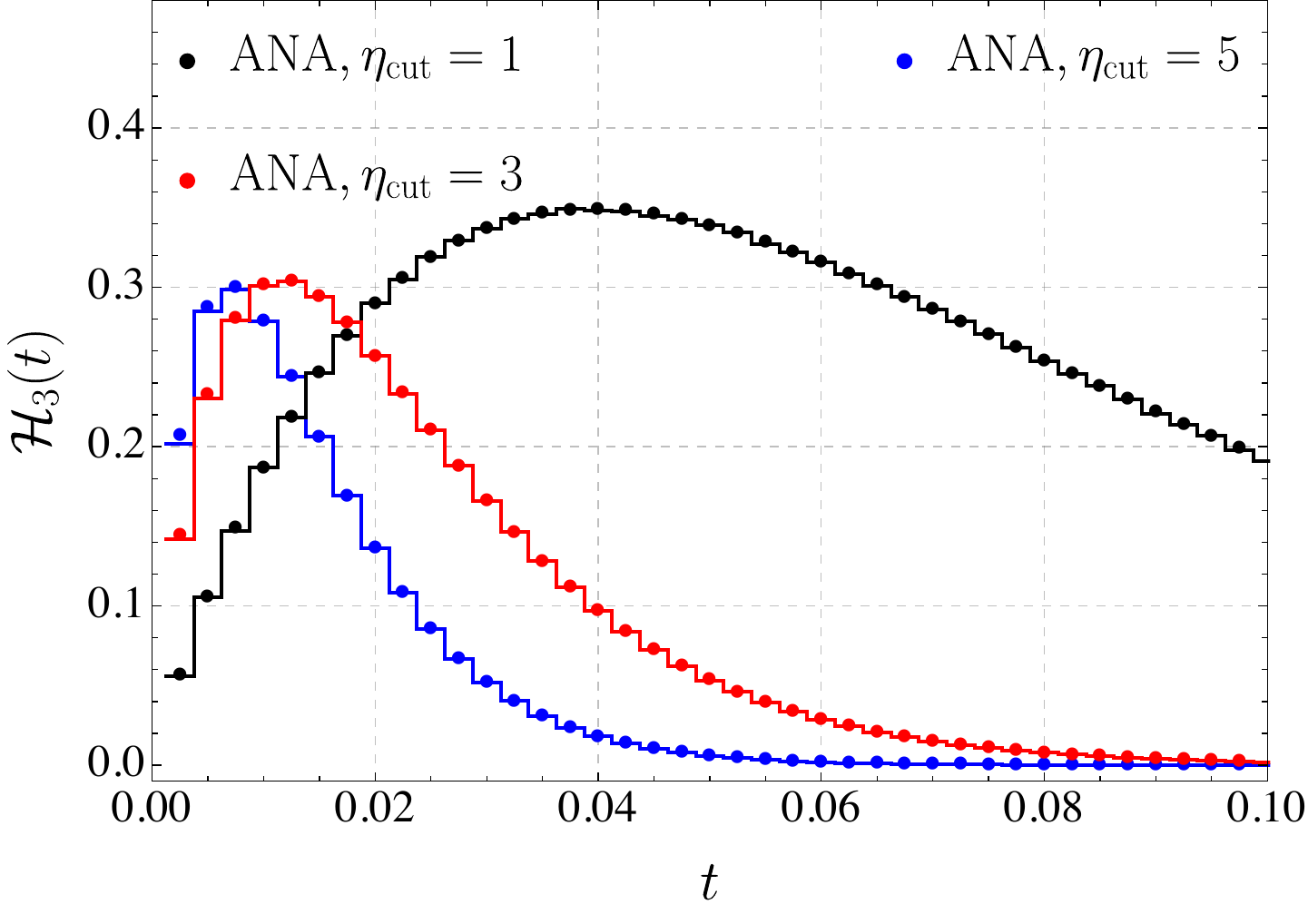}
 \hspace{.5cm}
 \includegraphics[width=0.47\textwidth]{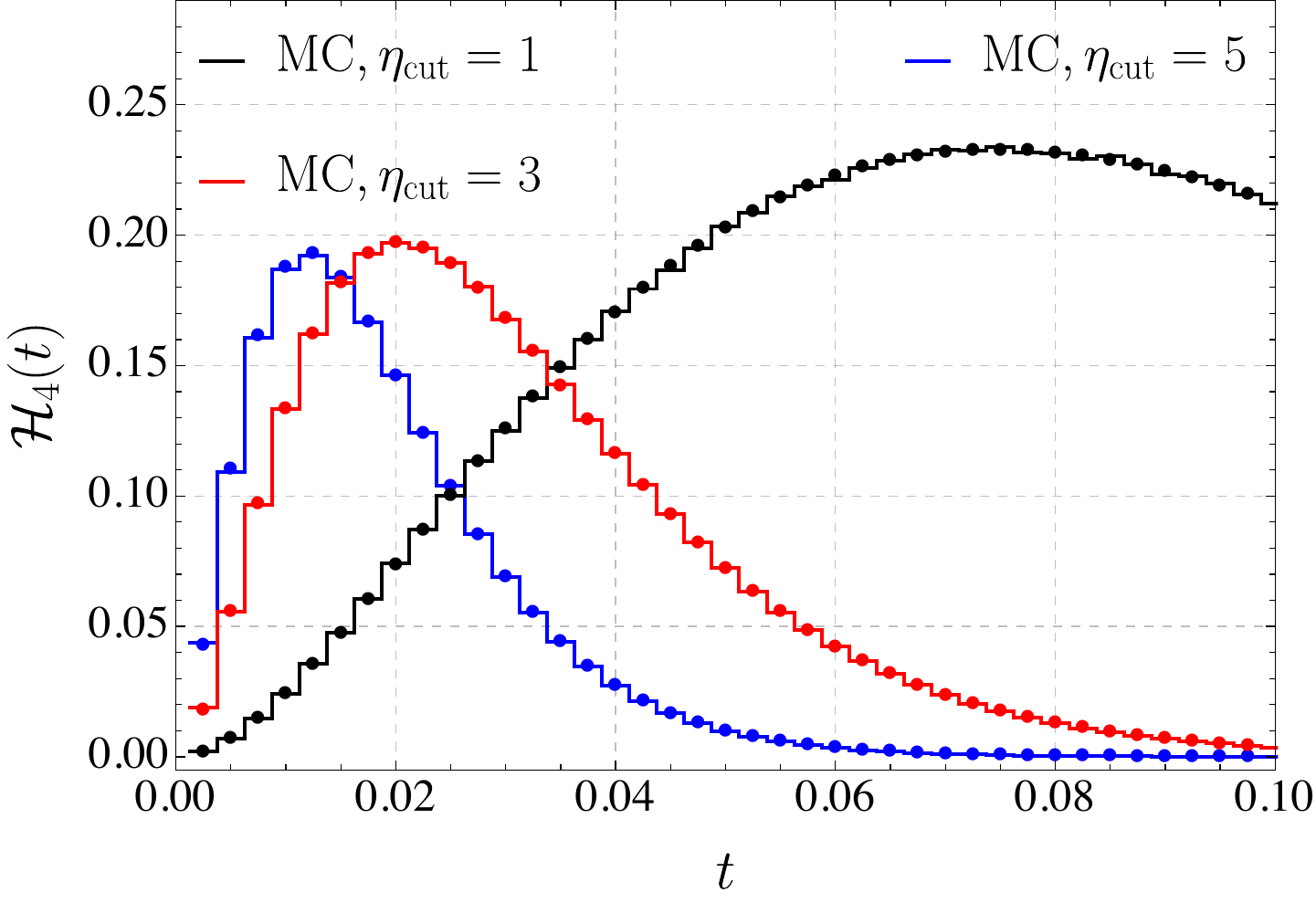}
 \caption{Numerical comparison between MC simulations and analytical calculations. The histograms represent MC simulations with different collinear cutoffs $\eta_{\rm cut}=1$ (black), $3$ (red) and $5$ (blue). The dots are from numerically integrating their analytical expressions.}
 \label{fig:Hm_fig}
\end{figure}

In order to check our MC simulation step by step, we can calculate  $\mathcal{H}_m$ directly from its definition, and then compare with simulation results.  We show the results for  $\mathcal{H}_3$ and $\mathcal{H}_4$ in Figure \ref{fig:Hm_fig}. The histograms represent the simulation results while the dots are calculated directly. For simplicity we set the veto region to zero which means that we do not veto any radiation. We write the collinear cutoff in the form $\lambda^2 = 1 - \tanh \eta_{\rm cut}$ and choose different values of $\eta_{\rm cut}$. We observe excellent agreement between the numerical integration and the simulation results. As a second consistency check we have verified the unitarity of the shower, i.e. we ran the full shower with the veto region to zero and checked ${\sigma}_{\rm veto}(t)=\sigma_0$ within the numerical accuracy.

We will also compare our simulation algorithm to the one used by Dasgupta and Salam \cite{Dasgupta:2001sh}. As mentioned in Appendix \ref{sec:labcone}, they impose the collinear cutoff in the COM rather than the lab frame. Furthermore, instead of computing the cross section directly, they formulate a shower for the derivative $d\sigma_{\rm veto}/dt$. This form can be derived from the differential form (\ref{eq:diffhrd}) of the RG equation. Specifically, we have 
\begin{align}\label{eq:dsS}
-\frac{1}{\sigma_0}\frac{d}{d t} \sigma_{\rm veto} = &  \int_\Omega \bm{3}_{\rm out}   \Big [ V_{2} \, e^{-t V_{2}} \Big] \frac{ R_{12}^3}{V_{2}} \nno \\
&  +  \int_\Omega \bm{4}_{\rm out} \bm{3}_{\rm in}  \int_0^t dt' \left [ V_{2} \, e^{-t' V_{2}} \right] \frac{ R_{12}^3}{V_{2}} \left [ V_{3} \, e^{-(t-t') V_{3}} \right]  \frac{ R_{132}^4 }{V_{3}}  \nno \\
& + \int_\Omega \bm{5}_{\rm out} \, \bm{4}_{\rm in} \, \bm{3}_{\rm in}  \int_0^t dt' \int_0^{t'} dt'' \left [ V_{2} \, e^{-t'' V_{2}} \right]  \frac{ R_{12}^3}{V_{2}} \left [ V_{3} \, e^{-(t'-t'') V_{3}} \right]  \frac{ R_{13}^4}{V_{13}}  \nno \\
& \hspace{1cm} \times \left \{ \frac{V_{13}}{V_3}  \left [ V_{4}^{(1)} \, e^{-(t-t') V_{4}^{(1)} } \right] \frac{ R_{1432}^5 }{V_{4}^{(1)}}
+  \frac{V_{32}}{V_3} \left [ V_{4}^{(2)} \, e^{-(t-t') V_{4}^{(2)} } \right]  \frac{ R_{1342}^5 }{V_{4}^{(2)}}\right \} \nno \\
& + \cdots ,
\end{align}
with $ \int_\Omega \bm{l}_{\rm out} = \int \frac{d\Omega(n_l)}{4\pi} \Theta_{\rm out}(n_l)$ and the abbreviation  $R_{1 i_1 i_2 \cdots i_m 2}^l = R_{1 i_1}^l  + R_{i_1 i_2}^l  + \cdots + R_{i_m 2}^l $. Equation \eqref{eq:dsS} immediately translates into a shower algorithm. One starts with the original dipole at $t=0$ as before. Then, for any event $E$ one generates a time-step according to $\mathcal{P}_E$, selects a dipole of the event with probability $\frac{V_{ij}}{V_E}$, and inserts a new vector into the dipole, splitting it into two. This is repeated until the new vector lies outside the jets (inside the veto region) at which point the shower is terminated and the value of $t$ is inserted into the histogram. This is the shower used in \cite{Dasgupta:2001sh}. 

\begin{figure}[t]
 \centering
 \includegraphics[width=0.7\textwidth]{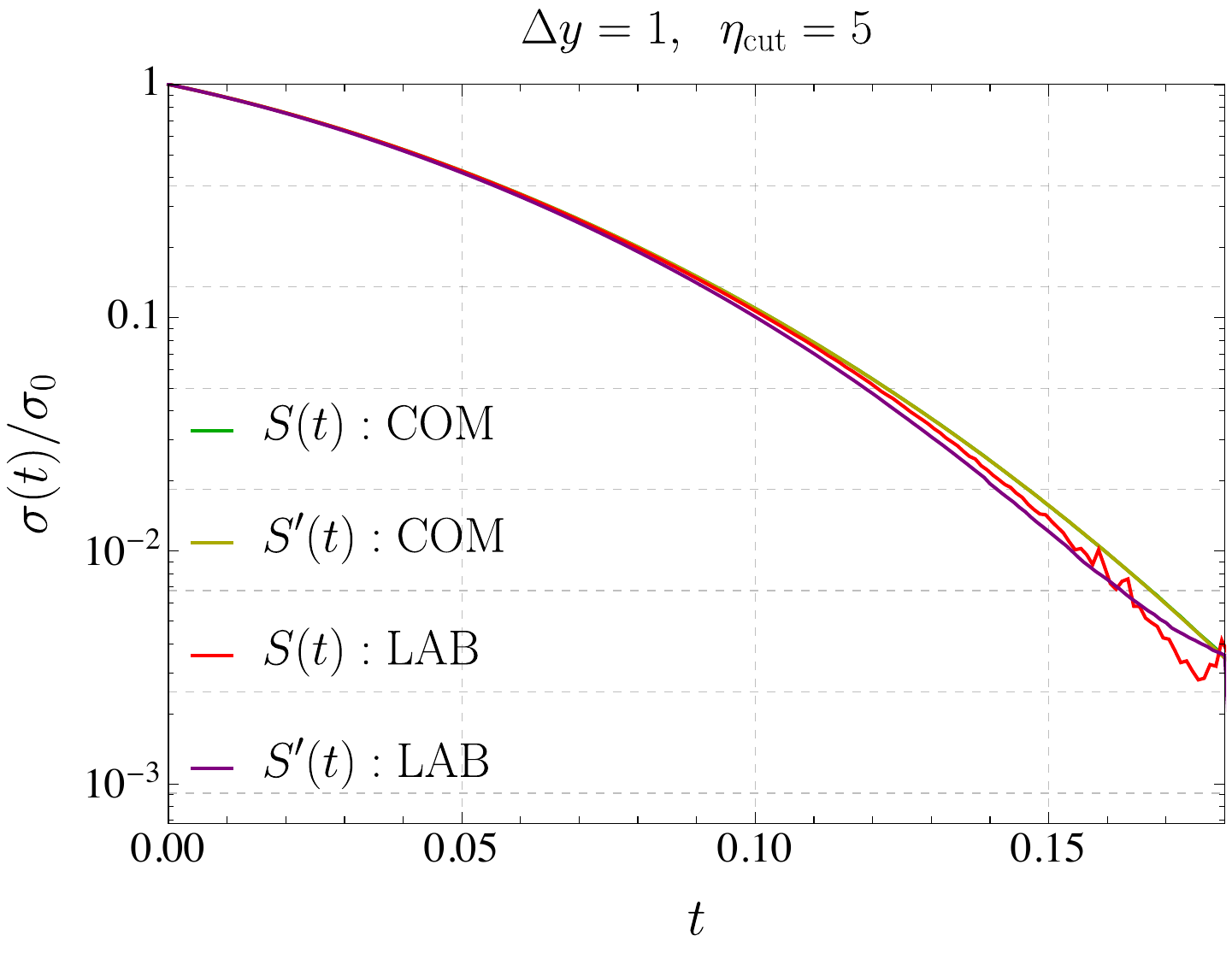}
 \caption{Numerical comparison between different simulation algorithms and collinear regularization methods (lab-cone versus center-of-mass cone). The curves labelled $S(t)$ are obtained from simulating the cross section, the ones labelled $S'(t)$ are obtained after simulating the derivative and integrating. The two COM curves are completely overlapping.}
 \label{fig:diffsk_fig}
\end{figure}
 
A numerical comparison of the different shower formulations and cutoff schemes is shown in Figure~\ref{fig:diffsk_fig}. Scheme $S(t)$ represents the algorithm we explain at the beginning of this appendix, and $S'(t)$ is the dipole shower of \cite{Dasgupta:2001sh} corresponding to the MC simulation of \eqref{eq:dsS}. For each algorithm, we show the two different ways to regularize the collinear divergence discussed in Appendix \ref{sec:labcone}. The curves labelled LAB apply the cutoff  \eqref{eq:labcut} in the lab frame, the ones labelled COM impose a rapidity cut in the center-of-mass frame of the emitting dipole. The two COM curves are nearly indistinguishable, while the curves in LAB cutoff scheme display small deviations beyond $t\gtrsim 0.1$. Comparing the different MC runs, we observe significant noise using the LAB cutoff at larger $t$. While the individual weights are close to one, larger-time entries involve many steps and we end up with some events with large weight which make the simulations noisy; conversely there are also many events with low weight which makes them inefficient. While any of the algorithms work well in the phenomenologically relevant region  $t< 0.1$, the COM scheme is clearly performing better at large $t$ and the algorithm simulating $S'(t)$ is especially well suited to get results at large $t$. A disadvantage of the $S'(t)$ scheme is that one needs to run it without any cutoff on $t$ in order to be able to reconstruct the function from the derivative. In contrast, one can restrict $t$ to the phenomenologically relevant region determined by the minimum value of $Q_0$ when directly generating the cross section. Also, when working with the cross section instead of the derivative, one can use the algorithm as an exclusive event generator and only impose the veto constraints at the end, after event generation.

\section{NLO expansion for isolated photon production\label{sec:NLOphoton}}

In this appendix we give analytical expressions for the lowest-order hard function and the NLO soft logarithm for isolated photon production at $e^+e^-$ colliders. 

If we expand to NLO, the factorization formula (\ref{masterFactorizationFormula}) truncates at $m=3$ since the hard functions scale as $\bm{\mathcal{H}}_{\gamma+n}\sim \alpha_s^{n-2}$. Expanding the ingredients in $\alpha_s$ and using that the lowest-order soft functions are trivial $\bm{\mathcal{S}}_{m}=\bm{1} +\mathcal{O}(\alpha_s)$, the cross section reads
\begin{align}\label{NLOexpIsoPhot}
\frac{d\sigma}{dx_\gamma}&=\langle\bm{\mathcal{H}}_{\gamma+2}^{(0)}\otimes \bm{1}\rangle+\frac{\alpha_s}{4\pi}\left[\langle \bm{\mathcal{H}}_{\gamma+2}^{(0)}\otimes\bm{\mathcal{S}}_{2}^{(1)}\rangle +\langle\bm{\mathcal{H}}_{\gamma+2}^{(1)}\otimes \bm{1}\rangle +  \langle \bm{\mathcal{H}}_{\gamma+3}^{(1)}\otimes\bm{1}\rangle\right]\,,
\end{align}
where the superscripts of $\bm{\mathcal{H}}_{\gamma+m}^{(n)}$ and $\bm{\mathcal{S}}_{m}^{(n)}$ indicate the order in $\alpha_s$. 

The hard function $\bm{\mathcal{H}}_{\gamma+2}^{(0)}$ describes the final state with one quark, one antiquark (with momenta $p_1$ and $p_2$) and one isolated photon (with momentum $p_\gamma$) in the final state. Using momentum conservation and introducing the variable $y_1 = (p_1 + p_\gamma)^2/(x_\gamma\, Q^2)$, we can write the LO hard function as
\begin{align}\label{HnloIsoPhot}
\bm{\mathcal{H}}_{\gamma+2}^{(0)}\left(y_1,Q, x_\gamma,\delta_0,\e\right)& =\sigma_0 \frac{\alpha\, Q_q^2}{2\pi}  \frac{e^{\gamma_E \e}}{\Gamma(1-\e)}\left(\frac{\mu}{Q}\right)^{2\e} \, \bar x_\gamma^{-\e} \,  x_\gamma^{-1-2\e}\nonumber\\
&\hspace{3cm}\times(y_1 \, \bar y_1)^{-1-\e} \, \left[ 2\, \bar x_\gamma + x_\gamma^2\,(y_1^2 + \bar y_1^2-\e)\right],
\end{align}
with $\bar x_\gamma = 1 - x_\gamma$, $\bar y_1 = 1 - y_1$, $Q_q$ is the charge of the quark flavour emitting the photon and $\sigma_0$ the associated Born cross section. Here we eliminated the bare fine-structure constant using $\alpha^0 = \tilde{\mu}^{2\e} \alpha =   \left[ e^{\gamma_E} \mu^2/(4\pi)\right]^\e \alpha$.  The Born cross section for the decay $\gamma^* \to q\bar q$ is given by
\begin{equation}
\sigma_0=N_c\,\alpha\, Q_q^2\, Q \frac{e^{\gamma_E \e}\, \Gamma(2-\e)}{\Gamma(2-2\e)}\left(\frac{\mu}{Q}\right)^{2\e}\,.
\end{equation}
The dependence on $y_1$ is the leftover angular integration after taking momentum conservation  into account  and enters the convolution with the soft function. The angular constraint, which enforces that the hard partons are outside the isolation cone, translates to an integration boundary in terms of $y_c = \frac{(1-\cos\delta_0)(1-x_\gamma)}{2-(1-\cos\delta_0)x_\gamma}$ as follows:
\begin{align}\label{ypara}
\bm{\mathcal{H}}_{\gamma+2}^{(0)}\left(\{n_1,n_2\},Q,E_\gamma,\delta_0\right)\otimes &\,\bm{\mathcal{S}}_{2}\left(\{n_1,n_2\},  \epsilon_\gamma E_\gamma,\delta_0\right)  \nno \\
& = \int_{y_c}^{1-y_c} d y_1\, \bm{\mathcal{H}}_{\gamma+2}^{(0)}\left(y_1,Q,x_\gamma,\delta_0\right)\bm{\mathcal{S}}_{2}\left(y_1, x_\gamma, \epsilon_\gamma,  \delta_0\right).
\end{align}

As the soft function is trivial at LO (first term on the right hand side of \eqref{NLOexpIsoPhot}),  we can immediately perform the integration over $y_1$, take the trace in color space and obtain the differential LO cross section as
\begin{align}
\frac{d\sigma^{\rm (0)}}{dx_\gamma}&=\int_{y_c}^{1-y_c} d y_1 \, \langle\bm{\mathcal{H}}_{\gamma + 2}^{(0)} \rangle \nonumber \\
&=\sigma_0\frac{ \alpha Q_q^2}{\pi} \left[\frac{2-2\,x_\gamma+x_\gamma ^2}{x_\gamma}\ln \left(\frac{1-y_c}{y_c}\right)-\left(1-2\,y_c\right)x_\gamma\right]\text{,}\label{eq:LOxsectionPhotonIsolation}
\end{align}	
 in agreement with the result in \cite{Kunszt:1992np}.

The second term $\langle \bm{\mathcal{H}}_{\gamma+2}^{(0)}\otimes\bm{\mathcal{S}}_{2}^{(1)}\rangle$ in \eqref{NLOexpIsoPhot} can be obtained by evaluating the soft function $\bm{\mathcal{S}}_{2}^{(1)}$ for one soft gluon inside the cone radiated off one of the Wilson lines along $\{n_1,n_2\}$, whose direction is parameterized by the variable $y_1$. The soft function reads 
\begin{align}\label{SnloIsoPhot}
\bm{\mathcal{S}}_2^{(1)}&= - 8\, C_F \frac{1}{\e} \left( \frac{\mu}{\epsilon_\gamma E_\gamma}\right)^{2\e} I(\e) \,\bm{1}\,,
\end{align}
with the angular integral 
 \begin{align}
 I(\e)= \int \frac{d\Omega_k}{4\pi} \frac{n_i \cdot n_j}{ n_i \cdot n_k \, n_k \cdot n_j} \theta( 1-\cos\delta_0 - n_k \cdot n_\gamma )\,. 
\end{align}
To extract the divergent part of the soft function, it is sufficient to evaluate the angular integral for $d=4$, where it can be rewritten in the form
\begin{equation}
I(0)= \int_{x_{\rm{min}}}^{x_{\rm{max}}} dx\left[1+\frac{2}{\pi}\arcsin \left(\frac{(1-2y_1)\sinh x -\xi\cosh x}{2\sqrt{\bar y_1y_1}}\right)\right]\,,
\end{equation}
after boosting to the center-of-mass frame of the emitting dipole.  We have introduced the abbreviation
\begin{align}
\xi &=\frac{(2-x_\gamma)\cos \delta_0+x_\gamma }{2-(1-\cos\delta_0)x_\gamma}\,,
\end{align}
and the integration boundaries which restrict the gluon to the inside of the isolation cone have the form
\begin{align}
x_{\rm{min}}&=\frac{1}{2}\ln\left[\frac{1+(1-2y_1)\xi-2\sqrt{\bar y_1y_1(1-\xi^2)}}{1-(1-2y_1)\xi+2\sqrt{\bar y_1y_1(1-\xi^2)}}\right]\,, \\
x_{\rm{max}}&=\frac{1}{2}\ln\left[\frac{1+(1-2y_1)\xi+2\sqrt{\bar y_1y_1(1-\xi^2)}}{1-(1-2y_1)\xi-2\sqrt{\bar y_1y_1(1-\xi^2)}}\right]\,.
\end{align}

The one-loop corrections of $\bm{\mathcal{H}}_{\gamma+2}^{(1)}$ and $\bm{\mathcal{H}}_{\gamma+3}^{(1)}$  could be extracted in numerical form using the results of \cite{Kunszt:1992np}. However, we are only interested in the logarithmic piece, so that the divergent part of the combination is sufficient. Since the cross section is finite, the divergence must be equal and opposite to the one in $\langle \bm{\mathcal{H}}_{\gamma+2}^{(0)}\otimes\bm{\mathcal{S}}_{2}^{(1)}\rangle$. Explicitly, we must find that it takes the form
\begin{align}
\langle\bm{\mathcal{H}}_{\gamma+2}^{(1)}\otimes \bm{1}\rangle +  \langle \bm{\mathcal{H}}_{\gamma+3}^{(1)}\otimes\bm{1}\rangle&= 8 \,  C_F  \frac{1}{\e} \left(\frac{\mu}{E_\gamma}\right)^{2\e} \int_{y_c}^{1-y_c} d y_1   \, \langle\bm{\mathcal{H}}_{\gamma + 2}^{(0)} \rangle\, I(0) \,.
 \end{align}
Adding the one-loop ingredients, we then obtain the NLO logarithmic terms as
\begin{align}
\frac{d\sigma^{(1)}}{dx_\gamma}&= 16 \,  C_F  \ln (\epsilon_\gamma)\int_{y_c}^{1-y_c} d y_1   \, \langle\bm{\mathcal{H}}_{\gamma + 2}^{(0)} \rangle\, I(0)\,.
\end{align}

\section{Narrow-cone limit of photon isolation \label{sec:NarrowConePhoton}}

To verify the factorization theorem for narrow isolation cones \eqref{factorizationFormulaSmallCone}, we apply the method of regions to the integral which arises in the computation of the differential cross section at leading order \eqref{eq:LOxsectionPhotonIsolation}. To apply the method, we write \eqref{eq:LOxsectionPhotonIsolation} in the form
\begin{align}
	\frac{d\sigma^{\rm (0)}}{dx_\gamma}&=\sigma_0\frac{\alpha Q_q^2}{2\pi}  \frac{e^{\gamma_E \e}}{\Gamma(1-\e)} \left(\frac{\mu}{Q}\right)^{2\e} \, \bar x_\gamma^{-\e} \,  x_\gamma^{-1-2\e}\,\mathcal{ I} \,
\end{align}
with the dimensionally regularized angular integral
\begin{align}	
	\mathcal{ I}&=\int d y_1 \, (y_1 \, \bar y_1)^{-1-\e} \, \left[ 2\, \bar x_\gamma + x_\gamma^2\,(y_1^2 + \bar y_1^2-\e)\right]\theta(1-y_c-y_1)\theta(y_1-y_c).\label{eq:integralLOisolatedPhoton}
\end{align}
For a narrow cone we have $y_c \approx \delta_0^2\, \bar{x}_\gamma/4 \ll 1$. The expansion of the integral $\mathcal{ I}$ gets contributions from three regions of the integration variable $y_1$: the hard region $h$, where $y_1$ is large $y_c\ll y_1 \approx 1$; the region $c$, where the photon is emitted collinear to the quark ($y_c\approx y_1 \ll 1$); and finally the region $\bar{c}$, where the photon is emitted collinear to the antiquark ($y_c\approx \bar{y}_1=1-y_1 \ll 1$). By expanding the integrand in each region to leading power and evaluating the resulting integrals, we get
\begin{align}
	\mathcal{I}_{h}&=\int_0^1 d y_1 \, (y_1 \, \bar y_1)^{-1-\e} \, \left[ 2\, \bar x_\gamma + x_\gamma^2\,(y_1^2 + \bar y_1^2-\e)\right]=-\frac{2(2\,\bar x_\gamma+x_\gamma^2)}{\e} + \mathcal{O}(\e)
\end{align}
for the hard region, and for the collinear regions we have
\begin{align}
	\mathcal{ I}_{c}& =\mathcal{ I}_{\bar{c}}=  \int_{y_c}^{\infty} d y_1 \, y_1 ^{-1-\e} \left[ 2\, \bar x_\gamma + x_\gamma^2(1-\e)\right]
	 = \frac{y_c^{-\e}}{\e}  \left[ 2\, \bar x_\gamma + x_\gamma^2(1-\e)\right] \, ,
\end{align}
because $\mathcal{I}$ is symmetric under $y_1\leftrightarrow\bar y_1$. Adding up the different contributions, we obtain
\begin{align}
	\frac{d\sigma^{\rm (0)}}{dx_\gamma}&=\sigma_0 \frac{e^{\gamma_E \e}}{\Gamma(1-\e)}\frac{\alpha Q_q^2}{2\pi}\left(\frac{\mu}{Q}\right)^{2\e} \, \bar x_\gamma^{-\e} \,  x_\gamma^{-1-2\e} (\mathcal{ I}_{h}+\mathcal{ I}_{c}+\mathcal{ I}_{\bar{c}})\nonumber\\
	&=\sigma_0 \frac{\alpha Q_q^2}{\pi} \, \Bigg \{ P_{\gamma \leftarrow q}(x_\gamma)\left [ -\frac{1}{\e} +\ln\left( \bar{x}_\gamma x_\gamma^2\right)-\ln\frac{\mu^2}{Q^2}\right] +\nonumber\\
	&\hspace{2.1cm} P_{\gamma \leftarrow q}(x_\gamma)\left [ +\frac{1}{\e} -\ln\left( \bar{x}_\gamma x_\gamma^2\right)+\ln\frac{\mu^2}{y_c Q^2}\right]- x_\gamma \Bigg\}+\mathcal{O}(\epsilon) \,, \label{eq:narrowConeVerif}
\end{align}
where we show the hard and the collinear contributions separately in the second and third line. 
The divergences of the individual terms in \eqref{eq:narrowConeVerif} are proportional to the splitting function	
\begin{equation}
P_{\gamma \leftarrow q}(z)= \frac{1+(1-z)^2}{z}\,,
\end{equation}
confirming our earlier statement that the two parts renormalize in the same way as the fragmentation function. 
Adding up the two pieces one ends up with the final result
\begin{align}	
	 \frac{d\sigma^{\rm (0)}}{dx_\gamma} &=\sigma_0 \frac{\alpha Q_q^2}{\pi}\left[ P_{\gamma \leftarrow q}(x_\gamma) \ln\left(\frac{1}{y_c}\right)-x_\gamma \right].
\end{align} 
This agrees with the expansion of the full result \eqref{eq:LOxsectionPhotonIsolation} to leading power in $y_c$, verifying our region expansion.

The contributions of the different momentum regions to \eqref{eq:narrowConeVerif} are in one-to-one correspondence to terms in the factorization theorem, which at leading order reduces to
\begin{align}\label{eq:factLO}
	\frac{d\sigma^{\rm (0)}}{dx_\gamma}=\frac{\text{d}\sigma^{\rm incl.}_{\gamma+q+\bar{q}}}{\text{d}x_\gamma}+2 \, \sigma_0 \,\left\langle \bm{\mathcal{J}}_{\!\! q\rightarrow\gamma+q}\left(\{\underline{n}\}, \delta_0 \,E_\gamma,x_\gamma \right)\otimes \bm{1}\right\rangle\text{,}
\end{align}
where the factor of two in front of the second term accounts for the identical contribution from the anti-quark. The hard region is the first term in \eqref{eq:factLO} and corresponds the cross section without isolation on the photon. The collinear region in the second line of \eqref{eq:narrowConeVerif} corresponds to the second term in \eqref{eq:factLO} which describes the production of a $q\bar{q}$ pair, followed by fragmentation of the quark. We thus confirm the factorization theorem \eqref{factorizationFormulaSmallCone} at LO.

\end{document}